\documentclass[aps,prd,showpacs,twocolumn,superscriptaddress,nofootinbib,preprintnumbers]{revtex4-1} 

\makeatletter
\def\p@subsection{}
\makeatother

\usepackage{graphicx,amsmath,amssymb,lipsum,bm}

\usepackage[usenames,dvipsnames]{xcolor}
\definecolor{xlinkcolor}{rgb}{0.7752941176470588, 0.22078431372549023, 0.2262745098039215}

\usepackage[colorlinks=true,citecolor=xlinkcolor,linkcolor=xlinkcolor,urlcolor=xlinkcolor, backref=false,pdfborder={0 0 0}]{hyperref}

\usepackage[normalem]{ulem}
\usepackage[utf8]{inputenc} 
\usepackage{mathtools}
\usepackage{aas_macros}
\usepackage{float}

\usepackage{multirow}
\usepackage{ulem}
\usepackage{diagbox}

\usepackage[dvipsnames]{xcolor}
\usepackage{mathrsfs}

\newcommand{\be}{\begin{equation}}
\newcommand{\ee}{\end{equation}}
\newcommand{\beqa}{\begin{eqnarray}}
\newcommand{\eeqa}{\end{eqnarray}}

\newcommand\G{\mathcal{G}_2}

\newcommand{\bseq}{\begin{subequations}}
\newcommand{\eseq}{\end{subequations}}

\renewcommand{\ln}{\mathop{\rm ln}\nolimits}

\newcommand{\ld}{\Lambda{\rm CDM}}

\newcommand{\wa}{w_0w_a{\rm CDM}}
\newcommand{\bao}{{\rm BAO}}
\newcommand{\cmb}{{\rm CMB}}

\newcommand{\sn}{{\rm SN}}

\def\gsim{\raise0.3ex\hbox{$\;>$\kern-0.75em\raise-1.1ex\hbox{$\sim\;$}}}
\def\lsim{\raise0.3ex\hbox{$\;<$\kern-0.75em\raise-1.1ex\hbox{$\sim\;$}}}

\def\beqn#1{\begin{equation}\label{#1}}
\def\eeqn{\end{equation}}

\def\beqa#1{\begin{eqnarray}\label{#1}}
\def\eeqa{\end{eqnarray}}

\newcommand{\Mpc}{\text{Mpc}}
\def\hMpc{h{\text{Mpc}}^{-1}}

\def\Z2{$\mathcal{Z_2}$}

\defcitealias{desi1}{Paper 1}
\defcitealias{desi2}{Paper 2}
\defcitealias{desi3}{Paper 3}
\defcitealias{desi4}{Paper 4}

\usepackage{xspace}
\newcommand{\paperone}{\citetalias{desi1}\xspace}
\newcommand{\papertwo}{\citetalias{desi2}\xspace}
\newcommand{\paperthree}{\citetalias{desi3}\xspace}
\newcommand{\paperfour}{\citetalias{desi4}\xspace}

\definecolor{darkgreen}{RGB}{0,120,0}
\newcommand{\resub}[1]{{\color{black}{#1}}}

\newcommand {\ignore}[1]{}

\begin{document}

\preprint{MIT-CTP/6007}

\title{{\Large Reanalyzing DESI DR1:}\\
5. Cosmological Constraints with Simulation-Based Priors
}

\author{Anton~Chudaykin}
\email{anton.chudaykin@unige.ch}
\affiliation{D\'epartement de Physique Th\'eorique and Center for Astroparticle Physics,\\
Universit\'e de Gen\`eve, 24 quai Ernest  Ansermet, 1211 Gen\`eve 4, Switzerland}
\author{Mikhail M.~Ivanov}
\email{ivanov99@mit.edu}
\affiliation{Center for Theoretical Physics -- a Leinweber Institute, Massachusetts Institute of Technology, 
Cambridge, MA 02139, USA} 
 \affiliation{The NSF AI Institute for Artificial Intelligence and Fundamental Interactions, Cambridge, MA 02139, USA}
\author{Oliver~H.\,E.~Philcox}
\email{ohep2@cantab.ac.uk}
\affiliation{Leinweber Institute for Theoretical Physics at Stanford, 382 Via Pueblo, Stanford, CA 94305, USA}
\affiliation{Kavli Institute for Particle Astrophysics and Cosmology, 382 Via Pueblo, Stanford, CA 94305, USA}

\begin{abstract} 
\noindent We analyze the public DESI full-shape clustering data 
using simulation-based priors (SBPs). Our priors are obtained by fitting normalizing flows to the distribution of EFT parameters measured from field-level simulations, themselves generated using tailored halo occupation distribution (HOD) models for each tracer. Incorporating SBPs in a power spectrum analysis significantly enhances $\Lambda$CDM cosmological parameter constraints; in combination with BAO information from DESI DR2 and a BBN prior on the baryon density, we find the matter density parameter $\Omega_m=0.2987\pm 0.0066$, the Hubble constant $H_0=68.80\pm 0.35\,\mathrm{km}\,\mathrm{s}^{-1}\mathrm{Mpc}^{-1}$, and the mass fluctuation amplitude $\sigma_8 = 0.766\pm 0.015$ (or the lensing parameter $S_8=0.764\pm 0.018$), which are $1\%$, $40\%$ and $50\%$ stronger than the baseline results, though with a notable downwards shift in $\sigma_8$, \resub{driven by the quasar HOD assumptions}. The SBPs also have a significant impact in extended models, with the dark energy figure-of-merit improving by $70\%$ ($20\%$) in a $w_0w_a$CDM analysis when combining with the CMB (and supernovae). In the SBP analysis, we do not find statistically significant evidence for dynamical dark energy: the equation of state parameters are consistent with a cosmological constant within $2.2\sigma$ ($1.4\sigma$) in analyses without (with) supernovae. The neutrino mass constraints are also enhanced, with the $95\%$ limits $M_\nu<0.073\,\mathrm{eV}$ and $M_\nu<0.090\,\mathrm{eV}$ in $\Lambda$CDM and $w_0w_a$CDM respectively. The latter is the strongest constraint obtained to date and reinforces the preference for the normal neutrino mass hierarchy, regardless of the background dynamics. While our results are sensitive to HOD modeling assumptions, they clearly demonstrate that the inclusion of small-scale information can significantly sharpen cosmological parameter constraints.
\end{abstract}

\maketitle

\section{Introduction}

\noindent Experimental probes of the large-scale structure (LSS) of the Universe have begun delivering 
percent-precision clustering statistics 
that cover progressively increasing 
cosmological volumes. Among them, 
the Dark Energy Spectroscopic Instrument (DESI) is in the process of creating the most detailed three-dimensional map of our Universe to date, which allows for high-precision tests of the standard cosmological model through
baryon acoustic oscillations, redshift-space distortions, and the full-shape of the galaxy clustering observables. 
The latter can be robustly explored using tools developed within the Effective Field Theory program~\cite{Baumann:2010tm,Carrasco:2012cv,Ivanov:2022mrd}\footnote{See also~\cite{Assassi:2014fva,Mirbabayi:2014zca,Senatore:2014eva,Senatore:2014vja,Senatore:2014via,Baldauf:2015xfa,Vlah:2015sea,Vlah:2015zda,Vlah:2016bcl,Vlah:2018ygt,Blas:2015qsi,Blas:2016sfa,Ivanov:2018gjr,Vasudevan:2019ewf,Chen:2020fxs,Chen:2020zjt} for important theoretical efforts in building the consistent EFT description of galaxies.}, as originally demonstrated
in~\cite{Ivanov:2019pdj,DAmico:2019fhj,Philcox:2020vvt,Philcox:2021kcw,Chen:2021wdi,Chen:2022jzq,Chen:2024vuf}. The EFT-based  full-shape 
analyses of DESI clustering information both within 
the collaboration~\cite{DESI:2024hhd}
and outside it \citep{desi1,desi2,desi3,desi4} have already produced some of the most competitive constraints on cosmological parameters to date, including for the late-time matter density, the amplitude of mass fluctuations, and the Hubble constant, as well as strong constraints on neutrino masses and dark energy dynamics.

Extracting the fundamental information 
from galaxy clustering is, however, limited by our ability to model the relationship between galaxies and dark matter. On large scales this relationship is described by the EFT bias
expansion~\cite{Assassi:2014fva,Senatore:2014eva,Mirbabayi:2014zca,Desjacques:2016bnm}, whilst on small scales 
it is typically modeled by means of 
simulations and empirical galaxy-halo
connection models (see~\cite{Wechsler:2018pic} for a review). In order to be 
conservative, the EFT-based full-shape analyses typically marginalize over EFT parameters
within large agnostic priors (see e.g.~\cite{Ivanov:2019pdj,Chudaykin:2020aoj,Chen:2021wdi,Maus:2024dzi}), which 
reflect our ignorance about the 
galaxy formation physics. 

A promising way beyond this conservative analysis is to incorporate advances in the numerical modeling of small-scale structure by calibrating EFT parameters with priors extracted from simulations~\cite{Ivanov:2024hgq,Ivanov:2024xgb,Ivanov:2024dgv,Cabass:2024wob,Akitsu:2024lyt,Ivanov:2025qie,Zhang:2025sfk,Chen:2025jnr,Zhang:2024thl} (see also an earlier work~\cite{Sullivan:2021sof} 
that put forward similar ideas in the context of the Halo-Zel'dovich model~\cite{Hand:2017ilm}
as well as later implementations of these ideas within the DESI collaboration~\cite{DESI:2025wzd}). 
Such simulation-based priors (SBPs)
allow one to efficiently encode non-linear information from N-body or hydrodynamical simulations into cosmological analyses, though are naturally limited by the modeling assumptions of the former. 
Recent work~\cite{Ivanov:2024xgb,Chen:2025jnr} has demonstrated that SBPs calibrated from modern halo occupation
distribution (HOD) models at the field level
can dramatically improve cosmological 
parameter inference by utilizing the information
that would be discarded in the usual
analyses.

In this paper we apply the simulation-based priors approach to a re-analysis of the public DESI full-shape clustering data~\citep{DESI:2025fxa}. 
Our galaxy SBPs are constructed using fairly general HOD models which account for secondary halo properties such as 
density-dependent and concentration-dependent assembly bias~\cite{Hearin:2015jnf,Cuesta-Lazaro:2023gbv,Yuan:2021izi,Yuan:2022rsc,Yuan:2023ezi,Rocher:2023zyh}, allowing us to capture a range of plausible 
galaxy-halo connections, but remain conservative with respect to HOD model theory systematics. Using this broad model space allows us to propagate realistic galaxy formation uncertainties into the cosmological parameter inference. 

This is the fifth paper of a series dedicated to re-analyzing the first-year DESI full-shape data. \paperone presented 
our custom likelihood including a pipeline 
to measure the power spectra and bispectra of DESI galaxies and quasars and accompanying $\Lambda$CDM constraints, 
\papertwo focused on the exploration of non-minimal cosmological models,
\paperthree delivered the strongest non-Gaussianity constraints from large-scale structure to date, while \paperfour derived the enhanced cosmological 
constraints on $\Lambda$CDM 
and its extensions by combining two- and three-dimensional DESI clustering data with CMB-lensing cross-correlations~\cite{Maus:2025rvz}.
This paper found the novel result that, in the absence of systematic effects, the full-combination of 
data favors the normal hierarchy\footnote{The normal and inverted hierarchies are two options for the ordering of the neutrino mass states depending on whether the first two or the last two of them have a small gap between them. The oscillation experiment bounds then translate into the lower bound on the total neutrino mass $0.06$ eV (normal hierarchy) or $0.1$ eV (inverted hierarchy)~\cite{Esteban:2024eli}.} of neutrino mass
states even in the 
$w_0w_a$CDM background, implying that the bound is robust to certain cosmological tensions and potential systematics in the CMB lensing data~\cite{Planck:2013nga,RoyChoudhury:2019hls,Craig:2024tky,Loverde:2024nfi,Green:2024xbb,Elbers:2024sha,Graham:2025fdt}.
In this paper we explore the 
role of the simulation-based priors on cosmological constraints based on the DESI galaxy power 
spectrum measurements
from \paperone.

The remainder of this paper is structured as follows. Section~\ref{sec:data}
outlines our data, 
theory predictions, 
analysis pipelines, 
and simulation-based priors, including the choice of HOD model.
Section~\ref{sec:res}
presents the main results,
including constraints 
on the standard cosmological 
parameters from the SBP-enhanced
DESI galaxy power spectrum, 
as well as limits on extended cosmological models in combination
with external CMB and Supernovae
datasets. We draw conclusions in Section~\ref{sec:concl}. 
Appendix~\ref{app:params} reports the full set of EFT parameter constraints, while Appendix~\ref{app:marg} quantifies the impact of projection effects when using SBPs. Appendix~\ref{app:bisp} discusses the role of the galaxy bispectrum in the SBP analysis, and finally, Appendix~\ref{app:Gauss} presents results obtained using Gaussian SBPs.

\section{Data and analysis}
\label{sec:data}

\subsection{Datasets}\label{sec:data1}

{\bf DESI DR1 full-shape statistics:} Our key dataset is the DESI DR1 power spectrum measured from the public data release
\citep{DESI:2025fxa}, which is supplemented in some analyses by the large-scale bispectrum. This is described in detail in \paperone; we summarize the main aspects below. The two- and three-point statistics are measured from six non-overlapping data chunks -- BGS, LRG1, LRG2, LRG3, ELG2, and QSO -- using quasi-optimal ‘unwindowed’ estimators implemented in the \texttt{PolyBin3D} code.
As discussed in \paperone, our power spectrum and bispectrum estimators and analysis pipeline correct for a number of systematic effects, including angular imaging systematics, radial integral constraints, wide-angle effects in the power spectrum, residual mask convolution and fiber collisions (treated deterministically for the power spectrum and stochastically for the bispectrum using the algorithm introduced in \paperone). 
We compute theoretical covariance matrices for both the power spectrum and bispectrum using the FFT-
based algorithms in \texttt{PolyBin3D}. These assume a Gaussian density field but incorporate the effects of survey geometry, fiber collisions, imaging systematics, and integral constraints. 

Our binning strategy and scale-cuts match those adopted in \paperone and \papertwo. For the power spectrum, we consider the range $k\in[0.02,0.20]\hMpc$ with a bin width $\delta k = 0.01\,\hMpc$, whilst for the bispectrum we restrict to $k\in[0.02,0.08]\,\hMpc$. We include three power spectrum multipoles ($\ell=0,2,4$) in our fiducial analysis, and the bispectrum monopole ($\ell=0$) in some extended 
analyses. Since our principal goal in this work is to analyze the power spectrum with simulation-based priors, we do not include the small-scale bispectrum or lensing cross-correlations in this work, nor the extended high-redshift quasar sample (which were featured in \paperthree and \paperfour). Our scale cuts have been previously validated against simulations
in~\cite{Ivanov:2019pdj,Nishimichi:2020tvu,Chudaykin:2020ghx,Ivanov:2021kcd,Ivanov:2021fbu,Chudaykin:2020hbf,Chudaykin:2022nru,Ivanov:2021zmi} and tested at the field level~\cite{Ivanov:2024hgq,Ivanov:2024xgb,Ivanov:2024dgv}.

{\bf DESI DR2 BAO:} We supplement our full-shape measurements with public baryon acoustic oscillation measurements from DESI DR2 \citep{DESI:2025zgx,DESI:2025zpo}. These include distance measurements from the six spectroscopic samples described above, as well as BAO from the Lyman-$\alpha$ forest and the LRG3 $\times$ ELG1 galaxy sample, with the latter omitted in the full-shape DR1 analysis due to systematic concerns. As argued in \paperone (and demonstrated explicitly in the appendices), the covariance between DR2 BAO and DR1 full-shape can be safely neglected due to the limited overlap in the galaxy sample and the weak correlation between the pre- and post-reconstructed statistics on small scales ~\cite{Philcox:2021kcw,Maus:2026wsb}.
We will collectively refer to these measurements as ``BAO''.

{\bf CMB:} We employ the high-$\ell$ \textsc{plik} TT, TE, and EE spectra, low-$\ell$ \textsc{SimAll} EE and low-$\ell$ \textsc{Commander} TT likelihoods from the official \textit{Planck} 2018 release~\cite{Planck:2018nkj}. 
In addition, we use measurements of the lensing potential auto-spectrum from Planck PR4 and ACT DR6 \cite{Carron:2022eyg,ACT:2023kun,ACT:2023dou}.
We will refer to these CMB measurements simply as ``CMB''.

{\bf Supernovae:} We additionally utilize Type Ia supernova data from the Pantheon+ sample \citep{Brout:2022vxf}, which includes 1550 spectroscopically confirmed SNe in the redshift range $0.001 < z < 2.26$. We employ the public likelihood from~\cite{Chudaykin:2025gdn} and analytically marginalize over the supernova absolute magnitude.
Hereafter, we will denote this supernova dataset as ``SN''.

\subsection{Cosmological models and parameters}\label{sec:data2}

\noindent In this work, we investigate three cosmological scenarios: the standard model ($\ld$), dynamical dark energy with equation-of-state $w(a) = w_0 + w_a(1 - a)$ ($\wa$), and non-minimal neutrino masses, described by $M_\nu\equiv\sum m_\nu$. 

Evolution in the $\ld$ model depends on six baseline parameters: $H_0,\omega_c\equiv \Omega_ch^2, \omega_b \equiv \Omega_bh^2, A_s, n_s,\tau$. When analyzing data combinations without the CMB, we fix $\tau$ to the CMB best-fit (though see \citep{Sailer:2025lxj}) and impose a wide prior on $n_s$, following the official DESI analysis~\cite{DESI:2024hhd}. We also fix the baryon density, $\omega_b$, to the mean of the BBN prior from \cite{Schoneberg:2024ifp}. Our results remain essentially unchanged if a CMB prior is instead placed on $\omega_b$, as the current LSS data are unable to constrain this parameter.
When using the CMB likelihoods, we additionally vary the optical depth $\tau$ and the baryon density $\omega_b$ with broad priors. If the total neutrino mass is fixed, we assume a single massive state with $M_\nu=0.06$ eV; if it is varied, we assume three degenerate mass eigenstates with the physical prior $M_\nu>0$. Finally, in the $\wa$ model, we sample the dark energy parameters, $w_0$ and $w_a$, imposing the priors specified in \papertwo.

We perform Markov Chain Monte Carlo (MCMC) analyses to sample from the posterior distributions using Metropolis-Hastings algorithm as implemented in the \textsc{Montepython} code~\cite{Audren:2012wb,Brinckmann:2018cvx}.
The plots and marginalized constraints are computed using the public \textsc{getdist} code \citep{Lewis:2019xzd}.\footnote{\href{https://getdist.readthedocs.io/en/latest/}{https://getdist.readthedocs.io/en/latest/}} We adopt a Gelman-Rubin~\cite{Gelman:1992zz} convergence criterion $|R-1|<0.02$ for all sampled parameters. For frequentist tests, we compute best-fit parameters by minimizing the likelihood with a Jeffreys prior, which is exactly equivalent to minimizing the full unmarginalized likelihood.

\subsection{Theoretical models}

\noindent Our analysis is based on the three-dimensional power spectrum multipoles and the large-scale bispectrum
monopole, which are described
with EFT to the one-loop and tree-level
order, respectively. 
The details of these computations
are presented in \paperone (see \paperfour for the extension of the latter to one-loop order). Schematically, 
the power spectrum predictions can be written as
\be 
P_\ell = P_{\ell}^{\rm tree}+P_{\ell}^{\rm 1-loop}+
P_{\ell}^{\rm ctr}+P_{\ell}^{\rm stoch}\,,
\ee 
where $P_{\ell}^{\rm tree}$ is the tree-level prediction based on the Kaiser linear theory
model \citep{Kaiser:1987qv}, $P_{\ell}^{\rm 1-loop}$ 
is the one-loop computation
in standard perturbation theory~\cite{Bernardeau:2001qr},
which depends on non-linear galaxy bias 
parameters $b_2,b_{\G},b_{\Gamma_3}$,
while $P_{\ell}^{\rm ctr}\propto c_\ell k^2 P_{\rm lin}(k)$ is the counterterm contribution that corrects the error in the 
standard perturbation theory (SPT) loop computation and accounts for the backreaction of small-scale effects on the large-scale distribution, including the 
fingers-of-God phenomenon~\cite{Jackson:2008yv}.
In addition, following~\cite{Ivanov:2019pdj,Nishimichi:2020tvu,Chudaykin:2020hbf} 
we include a single higher-order counterterm 
$\propto b_4 k^4 P_{\rm lin}(k)$
that captures higher order fingers-of-God
effects (with a characteristic angular dependence).  
Finally, $P_{\ell}^{\rm stoch}$
are multipole moments of the stochasticity
power spectrum, which read:
\be 
P_{\ell}^{\rm stoch}=\frac{2\ell+1}{2\bar n}\int_{-1}^1 d\mu\left[P_{\rm shot}+(a_0+a_2\mu^2)\left(\frac{k}{k_{\rm NL}}\right)^2\right] \,,
\ee 
where $\bar n$ is the galaxy number
density and $k_{\rm NL}=0.45~\hMpc$ is the 
normalization scale that roughly matches the 
non-linear scale of luminous red galaxies
in real space at $z\approx 0.5$. 

Our full model for the galaxy power spectrum is specified by eleven non-cosmological parameters:
\[
\{b_1,b_2,b_{\G},b_{\Gamma_3},c_0,c_2,c_4,b_4,P_{\rm shot},a_{0},a_{2}\}.
\]
As discussed below, we will test two choices of prior on these parameters: the conservative priors
of~\paperone and the simulation-based priors developed 
in~\cite{Ivanov:2024hgq,Ivanov:2024xgb}. 
The tree-level bispectrum model has
three more parameters: the mixed stochastic 
parameter $B_{\rm shot}$, the bispectrum stochasticity
$A_{\rm shot}$, and the fingers-of-God counterterm
$\tilde{c}_1$; we impose conservative priors on these parameters in all cases.

\subsection{EFT priors}
\noindent In this work, we explore two alternative sets of EFT priors. First, we adopt conservative Bayesian priors (CBPs), which have been employed in numerous EFT-based full-shape analyses (including \cite{Philcox:2021kcw,Ivanov:2021kcd,Ivanov:2019pdj,Philcox:2021kcw,Ivanov:2021kcd,Chudaykin:2024wlw}). 
These are motivated by the perturbative nature of the EFT expansion, specifically by the requirement that higher-order corrections should not exceed the leading-order (tree-level) contribution.
Although well-motivated physically, CBPs are relatively broad, leading to strong degeneracies among EFT parameters that limit the cosmological constraining power of full-shape data analyses. To alleviate this, we test simulation-based priors (SBPs) on all power spectrum EFT parameters informed by state-of-the-art decorated HOD models. By providing physically plausible and (fairly) general predictions for the galaxy–halo connection, SBPs are more informative than CBPs, though come with the inherent limitations of HOD modeling.
Practically, these utilize small-scale information to break parameter degeneracies (analogous to replacing the EFT-based analysis with an HOD emulator prescription) thereby enhancing cosmological inference.

To mitigate parameter projection effects, we sample the parameters $b_1\sigma_8(z)$, $b_2\sigma_8^2(z)$, $b_{\mathcal{G}_2}\sigma_8^2(z)$, following \cite{desi1} (see also \citep{Maus:2024dzi}).
For CBPs, we implement the analysis strategy from \papertwo.
In particular, we rescale the EFT parameters that
enter quadratically into the likelihood by the late-time fluctuation amplitude, $\sigma_8(z)$, and the Alcock–Paczynski (AP) amplitude~\cite{Alcock:1979mp}, based on how they appear in the theoretical model, and marginalize over those combinations analytically \citep[cf.,][]{Tsedrik:2025hmj,Paradiso:2024yqh}.
A validation test of the standard analysis pipeline employing conservative EFT priors is provided in Appendix A of \papertwo.
\resub{In our approach, we model massive neutrinos using the ``cold dark matter+baryons'' (CDM+b) fluid prescription, which is known to significantly suppress the amplitude of scale-dependent effects induced by light massive relics~\cite{Villaescusa-Navarro:2013pva,Castorina:2013wga,Costanzi:2013bha,LoVerde:2013lta}. Therefore, we do not include the neutrino-induced scale-dependent growth into the kernels; see the related discussion in \papertwo.}

For SBPs, we impose HOD-informed priors on the above parameter combinations involving EFT parameters and $\sigma_8(z)$ 
and explicitly sample these combinations in our MCMC chains. 
In Appendix~\ref{app:marg} we show that this rescaling of EFT parameters with $\sigma_8$ substantially reduces prior volume effects.  We do not apply the AP rescaling in the SBP case, since AP distortions were not modeled in the \texttt{AbacusSummit} simulations and are therefore not encoded in our simulation-based priors (see below).

As stated above, we do not include the bispectrum monopole in analyses when SBPs are included. Our motivation is that the SBPs themselves provide much tighter constraints on the EFT parameters than the DESI DR1 galaxy bispectrum. This is confirmed in Appendix~\ref{app:bisp}, which demonstrates that the parameter constraints remain largely unaffected when the
galaxy bispectrum is included in the SBP analysis. In contrast, when CBPs are adopted, the addition of higher-order statistics is necessary to obtain physical values of the EFT parameters, as discussed in \paperone. As such, we combine $P_\ell$ and $B_0$ for all analyses with conservative EFT priors.

\subsection{HOD models}\label{sec:hod}

\noindent In this section we describe the HOD models for various tracers that are used to produce priors on the EFT parameters.

\textbf{1. Luminous Red Galaxies (LRG) and Bright Galaxy Survey (BGS).}
The halo occupation distribution 
models for LRG and BGS are based on an extension of the 
classic seven-parameter Zheng et al.\,model~\cite{Zheng:2004id,Zheng:2007zg}, which posits:
\be
\label{eq:HODforLRG}
\begin{split}
& \langle N_c\rangle(M)=\frac{1}{2}\left[1+\text{Erf}\left(\frac{\log M-\log M_{\rm cut}}{\sqrt{2}\sigma}\right)\right]\,,\\
& \langle N_s\rangle(M)=\langle N_c\rangle(M)\left(\frac{M-\kappa M_{\rm cut}}{M_1}\right)^{\alpha}\,,
\end{split} 
\ee
where $\langle N_c\rangle$,
$\langle N_s\rangle$ are the average occupations of halos by 
centrals and satellites, respectively, as a function of the 
halo mass $M$, and $M_{\rm cut}$, $\sigma$, $\kappa$, $M_1$, $\alpha$ are the basic HOD parameters. 
In order to incorporate assembly bias, 
we treat $M_{\rm cut}$
and $M_{1}$ as functions
of the halo concentration $c$
and the local matter 
overdensity $\delta$~\cite{Hearin:2015jnf,Cuesta-Lazaro:2023gbv,Yuan:2021izi,Yuan:2022rsc,Yuan:2023ezi}.
The strength of the 
concentration and density-dependent assembly bias for centrals and satellites is captured by four additional parameters: 
$A_{\rm cen}, A_{\rm sat},B_{\rm cen}, B_{\rm sat}$.
We also utilize a satellite radial distance parameter $s$ to model the baryonic feedback~\cite{Yuan:2018qek}, and model the velocity bias following~\cite{Guo:2014iga,Yuan:2021izi}. 

To model redshift-space distortions, the line-of-sight
projection of the central galaxies'
velocity is scattered with random Gaussian fluctuations $\delta v$ whose variance is 
given by the velocity
dispersion $\sigma_{v}$
of dark matter particles
in a relevant halo. 
This fluctuation is then additionally
rescaled by the velocity bias coefficient $\alpha_c$ to yield
the following central galaxy velocity:
\be 
\begin{split}
& v_{\rm cen, z}=v_{L2,z}+\alpha_c \delta v(\sigma_{v})\,,
\end{split}
\ee 
where $v_{L2,z}$ is the 
subhalo velocity projected onto the line-of-sight. These satellite velocities are then shifted from the host particle velocity, $v_p$, according to the velocity-bias parameter $\alpha_s$:
\be 
\begin{split}
& v_{\rm sat, z}=v_{L2,z}+\alpha_s (v_{p,z}-v_{L2,z})\,.
\end{split}
\ee 
All in all, our HOD model for the LRG and BGS sample is specified by 12 HOD parameters.

To produce a sample of HOD models, 
we draw from the following flat and conservative
priors on the HOD parameters for BGS and LRG galaxies following \cite{Ivanov:2024xgb}:
\be
\begin{split}
& 
\log_{10} M_{\rm cut}\in [12,14]\,,\quad 
\log_{10} M_1\in [13,15]\,,\\
& \log \sigma \in [-3.5,1.0]\,,\quad  \alpha \in [0.5,1.5]\,,\quad \\
& \alpha_c \in [0,1]\,,
\quad 
\alpha_s \in [0,2]\,,
\quad s\in [-1,1]\,,
\quad  \kappa \in [0.0,1.5]\,,\\
& A_{\rm cen} \in [-1,1]\,,\quad 
A_{\rm sat} \in  [-1,1]\,,\\
& B_{\rm cen} \in [-1,1]\,,\quad 
B_{\rm sat} \in  [-1,1]\,.
\end{split}
\ee 
Rather than recomputing the priors, we utilize the 10500 HOD samples that have been previously used in the analysis of BOSS data in~\cite{Ivanov:2024xgb} (since our HOD model is the same). 
For each HOD parameter combination the former work generated a galaxy sample in a periodic box of the \texttt{AbacusSmall} suite,
and extracted the EFT parameters using the field-level EFT model
of~\cite{Schmittfull:2018yuk,Schmittfull:2020trd,Obuljen:2022cjo} following the transfer function
computations described in~\cite{Ivanov:2024xgb}.

Here, our samples are generated at a fixed 
redshift $z=0.5$ and for a fixed cosmological model of the \texttt{AbacusSummit small} suite. While our assumptions of fixed cosmology and redshift represent important limitations of our method, there is some evidence that variations of cosmology within
observationally allowed ranges
yield only small changes to the EFT parameters of HOD models compared to variation of the 
HOD parameters themselves~\cite{Ivanov:2024xgb,Ivanov:2025qie}.\footnote{See \citep{Zhang:2024thl,Zhang:2025sfk} for analyses including cosmology-dependence. Note that these analysis was carried out  
for a wider range of cosmological parameters but a more narrow range of HOD parameters than considered in \cite{Ivanov:2025qie}. 
} We also note that our work does not explore variation of the priors with respect to neutrino or dynamical dark energy cosmologies; since the primary effect of these contributions is to change the cosmological background and the clustering amplitude we similarly do not expect large changes to the HOD priors. \resub{This is consistent with 
 results of~\cite{LoVerde:2014rxa,Li:2024wco}, which found changes of the halo mass function due to non-Universality effects to be less than $10\%$ for massive neutrino and dynamical dark energy models within the ranges of their parameters probed by data. 
Potential variations of EFT parameters in a $\sim 10\%$ range due to these effects are negligibly smaller than those produced by the variation of the HOD parameters themselves, which are at the level of $\sim 100\%$~\cite{Ivanov:2025qie}.}

The DESI collaboration uses a similar version of the HOD function for the BGS sample which features improved behavior in the tails of the distribution~\cite{Smith:2023jqs,Findlay:2024oev,DESI:2025wzd} 
Since this change does not appreciably alter the EFT parameter distributions, we proceed with the simpler form~\eqref{eq:HODforLRG} in our analysis. 

\resub{We would like to stress that the EFT prior distributions
widen quite significantly when one switches from the vanilla HOD model to the decorated prescriptions~\cite{Ivanov:2025qie}. Hence, one may get much more constraining result in the case of the basic HOD models, but these results would also be more prone to biases due to model 
miss-specification. 
We use the decorated models in this work in order
to reduce this uncertainty
and in order for our results to be more conservative. }

\vskip 4pt

\textbf{2. Emission line galaxies (ELG).} To model the higher-redshift emission line galaxies, we use the HOD-ELG model and priors specified in~\cite{DESI:2023ujh}.
This model, which is also known as the High Mass Quenched (HMQ) prescription,
has been developed in \cite{Alam:2019pwr,eBOSS:2020yql,Yuan:2021izi,Rocher:2023zyh}, and takes a different form to the LRG and BGS model due to the inherent differences in the galaxy population (which is dominated by small satellites in contrast to large centrals).
The central galaxy HOD is given by
\be 
\label{eq:hod_elg}
\begin{split}
\langle N_c\rangle(M) &=2A\phi(M)\Phi(\gamma M)\\
&+\frac{1}{2Q}\left(1+\text{Erf}\left(\frac{\log(M/M_{\rm cut})}{0.01}\right)\right)\,,
\end{split}
\ee 
where the quenching efficiency is set to 
$Q=100$ following~\cite{Yuan:2022rsc}, and we have introduced
\be
\begin{split}
&
\phi(x) = \mathcal{N}(\log M_{\rm cut},\sigma_M)\,,\\
&\Phi(x) = \frac{1}{2}\left[1+\text{Erf}\left(\frac{x}{\sqrt{2}}\right)\right]\,,\\
& A=\frac{p_{\rm max-1/Q}}{\max[2\phi(x)\Phi(\gamma x)]}\,.
\end{split}
\ee 
For the satellites, we use the same 
power-law HOD function as for LRGs, \eqref{eq:HODforLRG}, but do not include the $\langle
N_c\rangle$ factor.

We additionally include assembly bias, baryonic feedback, 
and velocity bias  using the same approach as for LRGs. 
We sample the ELG-HOD parameters from the following
uniform distributions: 
\be 
\begin{split}
& 
\log M_{\rm cut}\in [11.6,12.6]\,,\quad 
\log M_1\in [12.5,18]\,,\\
& \log \sigma \in [-3.5,1.0]\,,\quad  \alpha \in [0,1.2]\,,\quad \\
& 
\alpha_c \in [0,1]\,,
\quad 
\alpha_s \in [0,2]\,,
\quad s\in [-1,1]\,,
\quad  \kappa \in [0,10]\,,\\
& p_{\rm max}\in [0.05,1]\,,\quad \gamma \in [1,15] \\
& A_{\rm cen} \in [-1,1]\,,\quad 
A_{\rm sat} \in  [-1,1]\,,\\
& B_{\rm cen} \in [-1,1]\,,\quad 
B_{\rm sat} \in  [-1,1]\,.
\end{split}
\label{eq:elg-prior}
\ee 
To generate the priors, we use 10500 ELG-HOD samples
generated in~\cite{Ivanov:2024dgv}
at $z=1.1$ based on the \texttt{AbacusSummit small} suite.
\vskip 4pt

\textbf{3. Quasars (QSO).}
To model the quasar sample, we use the same seven-parameter HOD as for LRGs:
\be
\label{eq:HODforQSO}
\begin{split}
& \langle N_c\rangle(M)=\frac{1}{2}\left[1+\text{Erf}\left(\frac{\log M-\log M_{\rm cut}}{\sqrt{2}\sigma}\right)\right]\,,\\
& \langle N_s\rangle(M)=\left(\frac{M-\kappa M_{\rm cut}}{M_1}\right)^{\alpha}\,,
\end{split} 
\ee
with the priors motivated by those of~\cite{Yuan:2023ezi}:
\be
\begin{split}
& 
\log_{10} M_{\rm cut}\in [11.2,14]\,,\quad 
\log_{10} M_1\in [14,16]\,,\\
& \log_{10}\sigma \in [-2,0.5]\,,\quad  \alpha \in [0.3,2.0]\,,\quad  \kappa \in [0.3,1.5]\,,\\
& \alpha_c \in [0,2]\,,
\quad 
\alpha_s \in [0,1]\,,
\end{split}
\ee 
For this sample, we do not supplement the model with assembly bias or baryonic feedback parameters. This choice is made for two reasons; first, the validity of the standard ``decorated'' models for quasars has not yet been
thoroughly established, and second, 
the results of~\cite{Yuan:2023ezi} 
suggest that the modeling of the DESI DR1 QSO sample
does not require secondary properties.
This can be contrasted with LRG galaxies,
which exhibit strong evidence
for assembly bias~\cite{Cuesta-Lazaro:2023gbv,Ivanov:2024xgb}.

To generate the EFT parameters for QSO, we use the approach of~\cite{Ivanov:2025qie}, which analytically maps from HOD to EFT parameters, under the assumption that the \textit{halo} bias parameters depend on the mass and cosmology only through 
the peak height.\footnote{Whilst this assumption remains to be rigorously tested for QSO, 
it was found to be quite accurate
for LRGs~\cite{Ivanov:2025qie}.
In addition, 
it is unlikely to be important here since the quasar sample contains only limited statistical weight.} 
We generate samples of EFT parameters at $z=1.5$,
which matches the effective 
redshift of DESI DR1 QSO sample.

For all three samples, we model the 
marginal distribution 
of EFT parameters using 
normalizing flows following 
the approach introduced in \cite{Ivanov:2024hgq}.
The flow-based models are then 
applied as priors on EFT parameters
which are explicitly sampled
in our MCMC chains (unlike in the CBP analyses, which analytically marginalize over most parameters).

\section{Results}\label{sec:res}

\begin{figure*}[!t]
	\centering
	\includegraphics[width=0.65\textwidth]{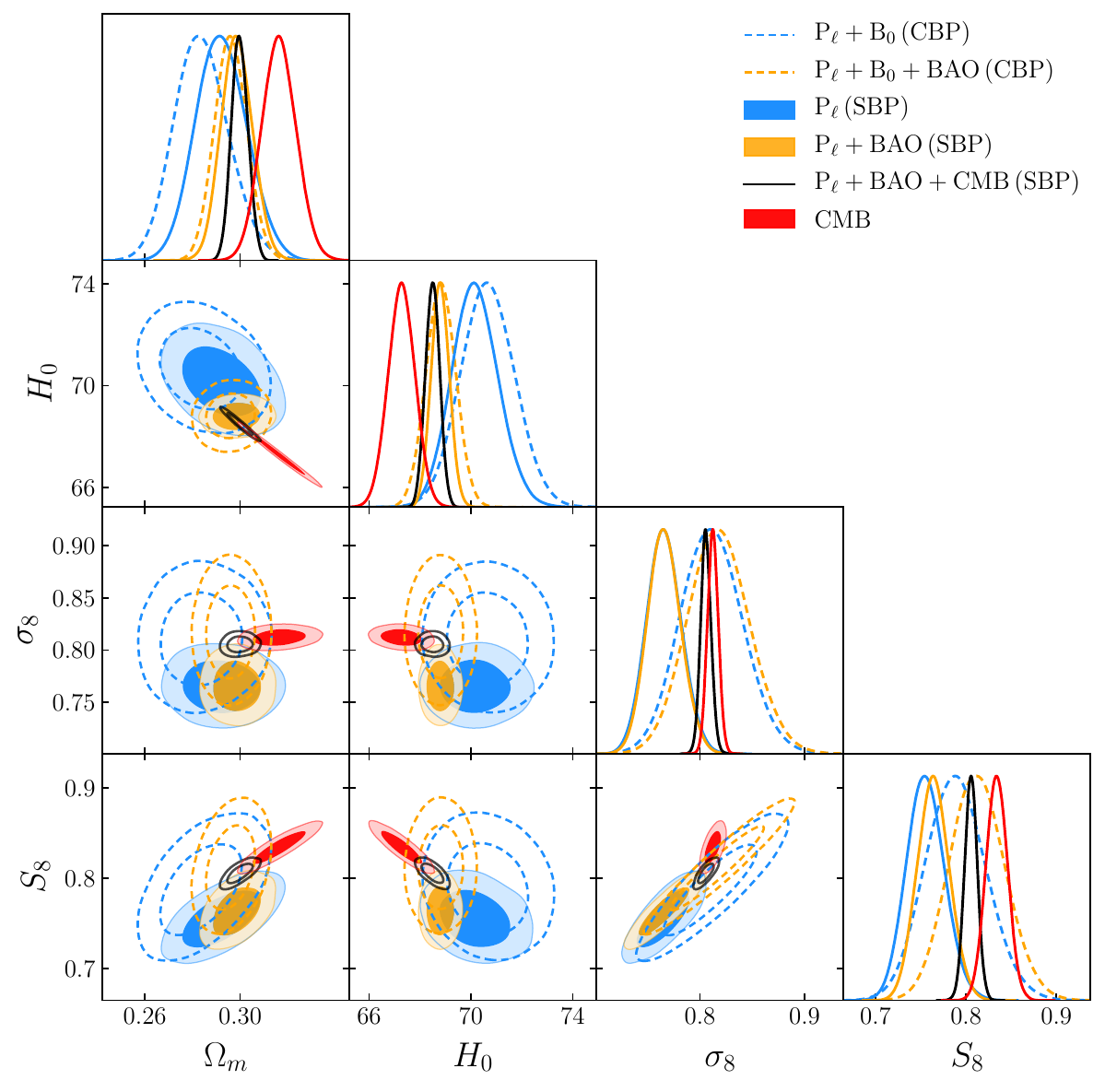}
	\caption{\textbf{$\Lambda$CDM constraints}: Posterior distributions for the $\ld$ parameters $\Omega$, $H_0$, $\sigma_8$ and $S_8$ obtained from different combinations of datasets: the DESI DR1 redshift-space power spectrum and bispectrum monopole ($P_\ell$ + $B_0$), expansion history measurements from DESI DR2 (BAO) and the {\it Planck} temperature and polarization anisotropy data, including lensing reconstruction from both the {\it Planck} \textsc{npipe} PR4 maps and ACT DR6 (CMB). Results are shown for conservative Bayesian priors (CBP, dashed) and simulation-based priors (SBP, solid). Since the addition of higher-order statistics does not alter the cosmological and EFT parameter constraints, we do not include the DESI DR1 galaxy bispectrum when including simulation-based priors; this is validated in Appendix~\ref{app:bisp}. The posteriors shrink significantly when SBPs are implemented, with a notable shift to lower values of $\sigma_8$ and $S_8$, which may indicate systematic effects or overly confident priors.
    The main parameter constraints derived from the CMB and DESI full-shape-plus-BAO datasets (using SBPs for the latter) are broadly consistent at the $2.8\sigma$ level, whilst mean values of $S_8$ differ between the two by $3.2\sigma$.
    \resub{The low values of $\sigma_8$ in the SBP analyses are driven by the more restrictive HOD models used for the quasar sample; replacing these with conservative EFT priors (but retaining HOD priors on other samples) increases $\sigma_8$ by approximately $1\sigma$, improving consistency with the baseline CBP result (see App.~\ref{app:qso} for details).}
    }
    \label{fig:ld}
\end{figure*}
\begin{table*}[!t]
    \centering
    %\resizebox{\linewidth}{!}{
    \begin{tabular}{lccccc}
    \toprule
    Dataset 
    & $\omega_{cdm}$ 
    & $\Omega_m$ 
    & $H_0$ 
    & $\sigma_8$ 
    & $S_8$  \\
    \hline
    %\enspace
    % SBP
    \textbf{SBP} &  &  &  & & \\
    $P_\ell$ (SBP)
    & $\enspace 0.1210_{-0.0055}^{+0.0049}\enspace$
    & $\enspace 0.292_{-0.011}^{+0.010}\enspace$ 
& $\enspace 70.18_{-0.94}^{+0.86}\enspace$ 
& $\enspace 0.766_{-0.016}^{+0.017}\enspace$ 
& $\enspace 0.756_{-0.022}^{+0.020}\enspace$ %final2
    \\
    $P_\ell+\bao$ (SBP)
    & $0.1185_{-0.0037}^{+0.0033}$ 
& $0.2987_{-0.0066}^{+0.0066}$ 
& $68.80_{-0.35}^{+0.35}$ 
& $0.766_{-0.016}^{+0.015}$ 
& $0.764_{-0.018}^{+0.018}$ %final2
    \\
    $P_\ell+\bao+\cmb$ (SBP)
    & $0.1175_{-0.0006}^{+0.0006}$ 
& $0.2999_{-0.0036}^{+0.0033}$ 
& $68.49_{-0.28}^{+0.28}$ 
& $0.8057_{-0.0049}^{+0.0049}$ 
& $0.8055_{-0.0069}^{+0.0070}$ %final2
    \\\hline
    % CBP
    \textbf{CBP} &  &  &  & \\
    $P_\ell$ 
    & $0.1122_{-0.0067}^{+0.0068}$ 
& $0.274_{-0.013}^{+0.012}$ 
& $70.22_{-1.06}^{+1.06}$ 
& $0.825_{-0.033}^{+0.033}$ 
& $0.787_{-0.036}^{+0.036}$ %final2
    \\
    $P_\ell+B_0$ 
    & $0.1189_{-0.0065}^{+0.0055}$ 
& $0.284_{-0.012}^{+0.010}$ 
& $70.67_{-1.05}^{+1.05}$ 
& $0.811_{-0.031}^{+0.028}$ 
& $0.789_{-0.035}^{+0.032}$ %final2
    \\
    $P_\ell+B_0+\bao$ 
    & $0.1174_{-0.0039}^{+0.0039}$ 
& $0.2961_{-0.0067}^{+0.0067}$ 
& $68.82_{-0.58}^{+0.58}$ 
& $0.818_{-0.029}^{+0.029}$ 
& $0.813_{-0.031}^{+0.031}$ %final2
    \\
    $P_\ell+B_0+\bao+\cmb$ 
    & $0.1172_{-0.0006}^{+0.0006}$ 
& $0.2983_{-0.0034}^{+0.0035}$ 
& $68.62_{-0.28}^{+0.28}$ 
& $0.8091_{-0.0056}^{+0.0051}$ 
& $0.8068_{-0.0074}^{+0.0075}$ %final2
    \\\hline
    $\cmb$ 
    & $0.1202_{-0.0012}^{+0.0012}$ 
& $0.3164_{-0.0073}^{+0.0072}$ 
& $67.28_{-0.53}^{+0.53}$ 
& $0.8123_{-0.0051}^{+0.0052}$ 
& $0.834_{-0.012}^{+0.012}$ %final2
    \\\hline
    \end{tabular}
    %}
    \caption{\textbf{$\bm{\Lambda}$\textbf{CDM} constraints}: Mean and 68\% confidence intervals on $\ld$ cosmological parameters obtained from various combinations of DESI and CMB data, adopting conservative Bayesian priors (CBP) and simulation-based priors (SBP) in the EFT-based full-shape analysis. Two-dimensional posteriors are shown in Fig.~\ref{fig:ld}.
    }
    \label{tab:ld}
\end{table*}
\begin{figure*}
\includegraphics[width=1\textwidth]{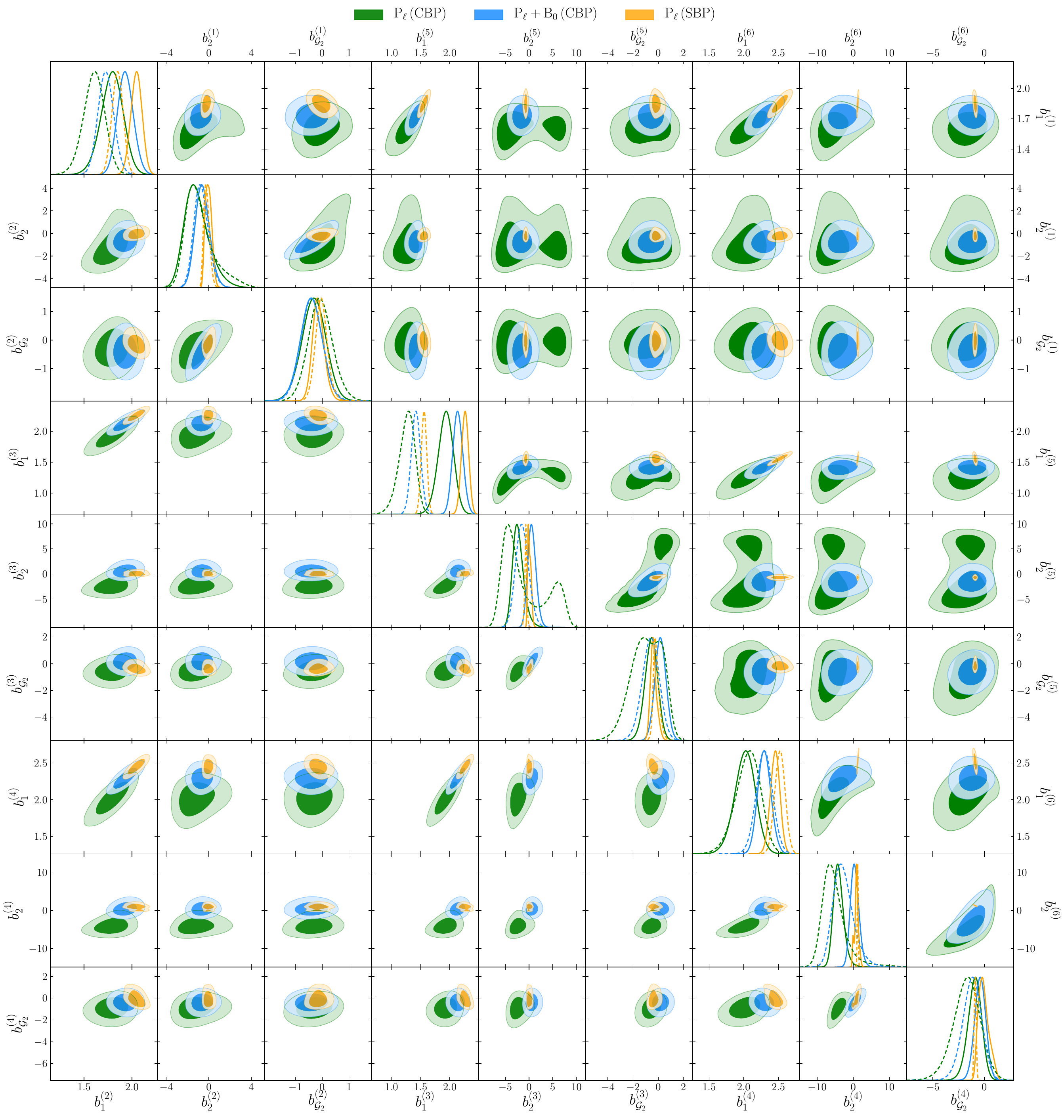}
\caption{\textbf{Bias parameters}: Posteriors on linear and quadratic bias parameters from $P_\ell$ and $P_\ell+B_0$ analyses using conservative Bayesian priors (CBP), and from $P_\ell$ analyses using simulation-based priors (SBP). The lower triangles show the bias parameters for the LRG1, LRG2, LRG3 galaxy samples, while the upper triangle presents those for BGS, ELG and QSO samples; these are represented by solid and dashed lines in the diagonal plots respectively. Superscripts (1)-(6) correspond to BGS, LRG1, LRG2, LRG3, ELG2, and QSO samples, respectively. 
Incorporating SBPs dramatically shrinks the posterior volume and pulls the bias parameters into closer agreement with the results of the standard $P_\ell+B_\ell$ analysis adopting CBPs. We note that the SBPs of this work apply also to the counterterm and stochasticity parameters, which are not shown above.
}
\label{fig:bias}
\end{figure*}

\subsection{\texorpdfstring{$\Lambda$CDM constraints}{LCDM constraints}}
\noindent Fig.~\ref{fig:ld} presents the main cosmological results for the $\ld$ model using the DESI clustering data combined with priors on $\omega_b$ and $n_s$, and optionally the CMB dataset. One-dimensional marginalized constraints given in Tab.~\ref{tab:ld}. 
Fig.~\ref{fig:bias} displays constraints on the linear and quadratic bias parameters for the six DESI data chunks for different analysis choices.

We first consider results obtained without the CMB data.
From Fig.~\ref{fig:ld}, it is clear that the addition of SBPs significantly increases the precision of cosmological parameter constraints compared to the case with conservative EFT priors, despite the inclusion of the bispectrum in the latter analysis.  
In particular, SBPs reduce the uncertainty on $\sigma_8$ by a factor of two in both the $P_\ell$ and $P_\ell+\bao$ analyses. Interestingly, this is accompanied by a downwards shift in $\sigma_8$ by \resub{$1.5\sigma_{\rm CBP}$ ($1.8\sigma_{\rm CBP}$)} 
%\antont{I replaced here $\sigma_{\rm SBP}$ with $\sigma_{\rm CBP}$, as the difference in $\sigma_{\rm SBP}$ is much larger. The difference is $1.8\sigma$ instead of $2\sigma$.} 
in the analysis without (with) BAO data, respectively, though the SBP and CBP final results remain consistent given the larger errorbars of the CBP analysis.
Moreover, the $H_0$ posteriors shrink by $15\%$ and $40\%$ in the $P_\ell$ and $P_\ell+\bao$ analyses, respectively.
We note that in the absence of BAO,
the improvements on $H_0$ are fairly muted, which is consistent 
with the notion that the physical distance scale information in the power spectrum is not substantially impacted by EFT parameter marginalization. We attribute a better $H_0$ inference 
in the combined SBP analysis to the degeneracy breaking imparted by the BAO dataset, which drastically changes the correlations in the $H_0-\Omega_m$ plane, see Fig.~\ref{fig:ld}.

\resub{
As mentioned above, the SBP analyses predict a somewhat lower value of $\sigma_8$ compared to those using conservative EFT priors.
This is attributed to the different EFT parameter values inferred in the two analyses.
In particular, the SBP analyses exhibit systematically higher linear biases: the mean value of $b_1$ is $\sim1.5\sigma$ higher across all DESI data chunks relative to the baseline CBP analysis (see Tab.~\ref{tab:EFT}).
Since $b_1$ and $\sigma_8$ are negatively correlated, a higher value of $b_1$ leads to a lower value of $\sigma_8$.}
%In Appendix~\ref{app:qso}, 

\resub{The dominant contributor to the $\sigma_8$ shift is the quasar sample, for which more restrictive HOD models (without assembly bias) were used to generate the SBPs.
This occurs for two reasons: (1) the QSO sample has both high redshift and large effective volume, giving enhanced sensitivity to the initial conditions; (2) SBPs provide informative constraints on the EFT parameters, significantly increasing the cosmological constraining power of this tracer. 
As shown in Appendix~\ref{app:qso}, adopting conservative EFT priors for the quasar sample and applying SBPs for the other tracers yields $\sigma_8=0.784_{-0.024}^{+0.024}$, corresponding to a shift of $1.2\sigma_{\rm CBP}$ relative to the baseline CBP result (compared to the $1.8\sigma_{\rm CBP}$ shift obtained in the main analysis of this work). 
This demonstrates that our results are sensitive to the choice of priors for the quasar sample.  
It is currently unclear whether this shift represents a physical effect, a statistical fluctuation or a systematic bias driven by overly optimistic EFT prior assumptions.
Clarifying this issue will require deriving quasar SBPs using more general HOD models, which we plan to investigate in future work.
%It is currently unclear whether this shift represents a physical effect, or is driven by overly optimistic EFT prior assumptions, systematic effects, or simply a statistical fluctuation.
%We plan to investigate this further in future work.
}

\resub{Our measurement is consistent with analyses of the BOSS data using SBPs~\cite{Ivanov:2024xgb}, as well as with the study based on $N$-body emulator-based models~\cite{Ibanez:2024uua}. These works favor moderately low values of the late-time fluctuation amplitude, $\sigma_8\simeq0.75$, in agreement with our findings.
They also incorporate small-scale information (learned from non-perturbative simulations), which effectively breaks degeneracies among EFT parameters and thus yield different, but potentially more robust, constraints on the EFT parameter space. 
As discussed above, the $b_1$ is negatively correlated with $\sigma_8$; therefore different values of the bias parameters imply different constraints on $\sigma_8$.
%This demonstrates that informative constraints on EFT parameters are essential for obtaining accurate inferences of cosmological parameters from the full-shape clustering statistics.
}

% \resub{Our constraints are dominated by the luminous red galaxy samples~\cite{DESI:2024jxi}.
% In Appendix~\ref{}, we present the results of the $P_\ell+BAO\,(SBP)$ analysis with the QSO sample omitted. The parameter constraints remain essentially unchanged, except for $\sigma_8$, whose error bar degrades by $30\%$, while its mean value remains essentially unchanged, $\sigma_8=0.764_{-0.023}^{+0.021}$.
% This indicates that the improvement in the $\sigma_8$ constraint is partially driven by the QSO sample, for which more restrictive HOD models (without assembly bias) were used to generate the SBPs. When the QSO sample is omitted, the $\sigma_8$ constraint in the SBP analysis is $25\%$ tighter than the baseline result using conservative EFT priors, compared to $50\%$ in the baseline SBP analysis including QSOs.
% It is therefore important to explore more flexible HOD models for QSO tracers in the SBP analysis. We leave this task for future work. 
%Notably, the preference for lower values of $\sigma_8$ remains robust to the addition of the QSO sample.
%}

The $\omega_{\rm cdm}$ constraint in the full-shape-only analysis tightens by $\approx 37\%$ when SBPs are added, but this improvement degrades to $10\%$ upon addition 
of the BAO data. This is justified by noting that the DR2 BAO data (including the high-redshift Lyman-$\alpha$ forest) provides much tighter constraints on $\Omega_m$ than the galaxy power spectrum shape, which dominates the overall constraints on $\omega_{\rm cdm}$.
In line with this argument, the $\Omega_m$ constraint from the joint full-shape and BAO analysis improves only marginally by $\lesssim 10\%$ when SBPs are employed.

Notably, the above SBP analyses produce CMB-independent cosmological parameter measurements that are as constraining as 
the ``kitchen sink'' analysis of \paperfour (though in slight tension, see Sec.~\ref{sec:concl}), which includes 
the one-loop bispectrum, cross-correlations with CMB lensing, 
and DESI's photometric sample. This is consistent 
with previous estimates of the significant loss
of information in the conservative EFT-based
analyses due to marginalization over EFT parameters~\cite{Wadekar:2020hax,Cabass:2022epm}.

The parameter constraints derived from DESI data alone with SBPs are broadly consistent with those from the CMB analysis (with maximal deviations of $2.8\sigma$ for the main cosmological parameters), which motivates combining the DESI and CMB datasets. As expected, the SBPs yield a more modest improvement in parameter constraints in this scenario, since the CMB already provides a tight constraint on the amplitude of the power spectrum. As in Tab.~\ref{tab:ld}, we find that the SBPs lead to improvements in $\sigma_8$ and $S_8$ of $6\%$ and $7\%$, respectively, but negligible changes to the $H_0$ and $\Omega_m$ posteriors.  

Next, we discuss the constraints on bias parameters. 
From Fig.~\ref{fig:bias}, we may draw two immediate conclusions.
First, using SBPs in place of CBPs significantly reduces the posterior widths of all bias parameters, with the final width considerably narrower than that obtained from a bispectrum analysis. 
Second, the constraints obtained with SBPs are broadly consistent with those in the baseline $P_\ell+B_0$ analysis using CBPs, which confirms the general robustness of our approach. 
Interestingly, SBP analyses constrain the linear bias at the $3\%$ level across all DESI samples; moreover, the quadratic bias parameters $b_2$ and $b_{\G}$ are determined with absolute uncertainties of $0.2-0.3$ for most samples. The main exceptions are LRG3, for which the uncertainties increase to about $0.5$, and QSO, which yields tight constraints with error-bars of approximately $0.1$ for both parameters (though we caution that the QSO sample uses the most restrictive HOD model).
% That said, certain values are in mild tension with the conservative results, such as the linear bias for LRG1, which induce slight shifts in cosmological parameter constraints\antont{Remove it.}. This difference between data- and simulation-derived (\textit{i.e.}\ $B_0$ versus SBP) constraints may also indicate subtle systematic effects in the DESI clustering data (including fiber assignment, angular foregrounds and beyond) that are usually absorbed into bias parameters.
The differences between data- and simulation-derived (\textit{i.e.}\ $B_0$ versus SBP) constraints can arise from subtle systematic effects in the DESI clustering data (including fiber assignment, angular foregrounds and beyond) that are usually absorbed into bias parameters.

The full EFT parameter constraints are provided in Tab.~\ref{tab:EFT}. 
SBPs significantly tighten several other EFT parameter constraints, including EFT
counterterms and stochastic parameters. Notably, the constraints on $b_{\Gamma_3}$ are weaker than those obtained in the standard $P_\ell+B_0$ analysis (for all tracers except QSO). 
This is because the decorated HOD models used to construct SBPs cannot constrain this parameter~\cite{Ivanov:2024hgq,Ivanov:2024xgb}; see Appendix~\ref{app:params} for further discussion.

\subsection{Dynamical dark energy}
\begin{figure*}
\includegraphics[width=0.99\columnwidth]{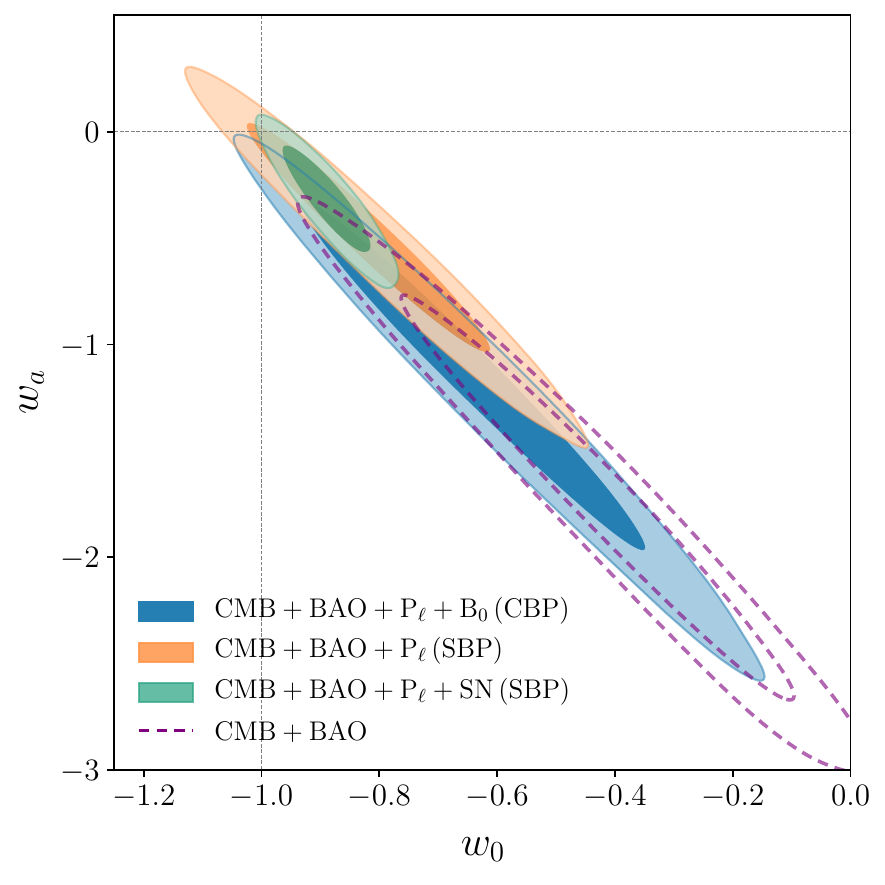}
\includegraphics[width=0.99\columnwidth]{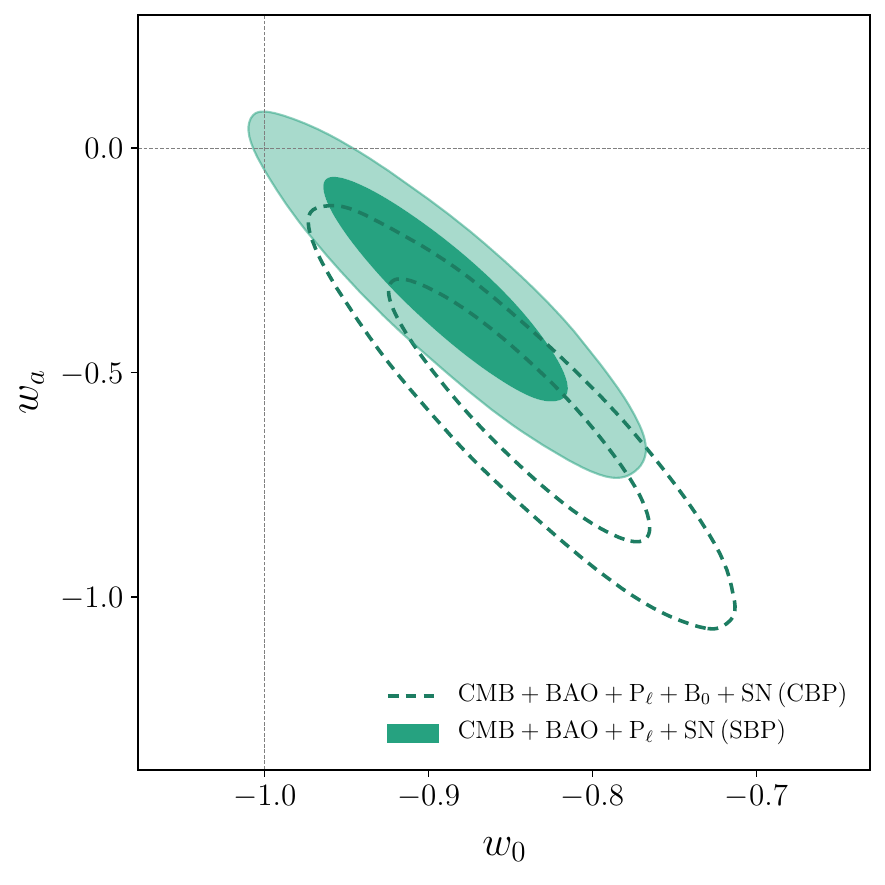}
\caption{\textbf{Dynamical dark energy}: Two-dimensional posterior distributions on the dynamical dark energy parameters for various combinations of datasets, as indicated by the captions. We show results with both conservative and simulation-based priors (CBPs \& SBPs). 
{\it Left panel:} The use of SBPs reduces the preference for dynamical dark energy from $2.8\sigma$ to $2.2\sigma$ without relying on supernova distance information (see Tab.~\ref{tab:chi2} for frequentist constraints)
{\it Right panel:} When supernova data are included, adopting SBPs alleviates the preference for a time-evolving dark energy equation of state in the conventional EFT-based analysis from $2.9\sigma$ to $1.4\sigma$.
}
\label{fig:w0wa}
\end{figure*}

\begin{figure}
\includegraphics[width=0.99\columnwidth]{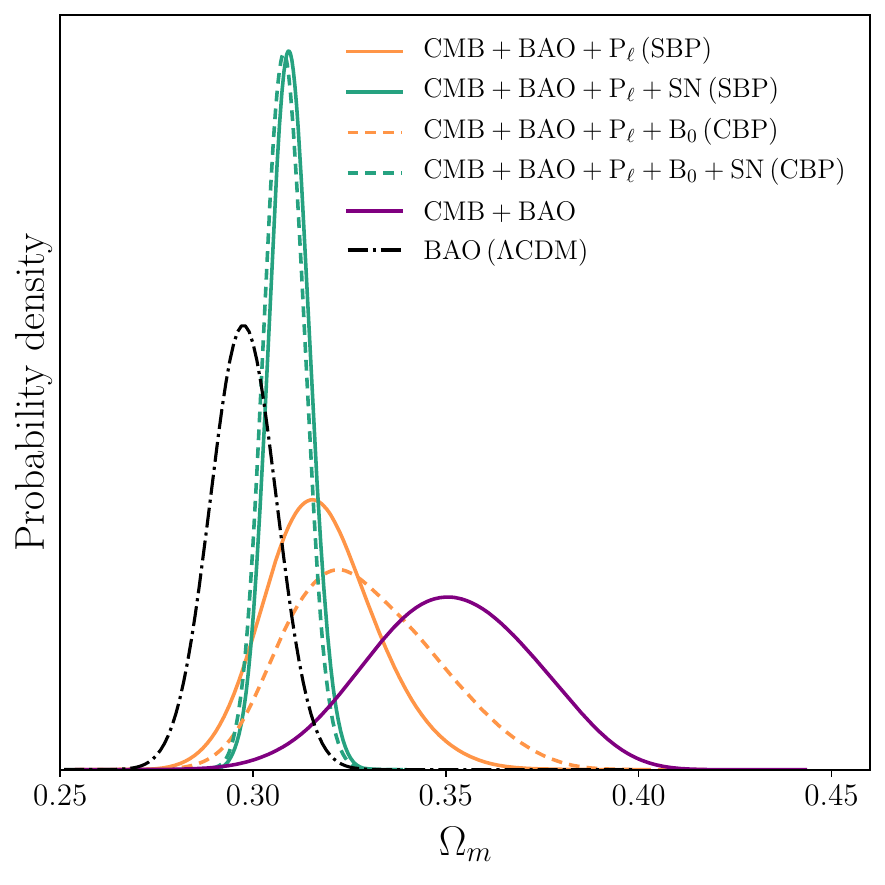}
\caption{\textbf{$\Omega_m$ constraints}: Marginalized one-dimensional posterior on $\Omega_m$, when fixing the background model to $\wa$. We also show the $\Omega_m$ posterior from the official DESI analysis of the BAO DR2 data in the $\ld$ model (black dot-dashed line). The addition of the DESI DR1 FS information to the CMB and BAO dataset tightens the $\Omega_m$ posterior by $11\%$ when using conservative Bayesian priors (CBP) and by $31\%$ when adopting simulation-based priors (SBP) and shifts the mean towards lower values in a closer agreement with the DESI-only $\ld$ result, without relying on supernova information.
}
\label{fig:w0wa_Om}
\end{figure}

\noindent Secondly, we constrain the $\wa$ model parameters by combining the DESI data with external information from CMB and 
supernovae. The one-dimensional marginalized parameter constraints are listed in Tab.~\ref{tab:w0wa}, with the results visualized in Fig.~\ref{fig:w0wa} (for $w_0-w_a$) and Fig.~\ref{fig:w0wa_Om} (for $\Omega_m$). Our main conclusion is clear: both with and without supernovae, we find a shift in the posterior distributions toward the 
cosmological constant when adding the SBPs, with the $\cmb+\bao+P_\ell$ dark energy contour consistent with $\ld$ within $68\%$\,CL.

Our results are in some tension with those obtained in the official DESI SBP study~\cite{DESI:2025wzd}, which found less improvement in the $(w_0,w_a)$ plane despite a less conservative HOD model. This arises since the former work calibrated the HOD priors using power spectrum measurements rather than directly from the simulation density fields (as in this work). This does not benefit from cosmic variance cancellation and results in comparative large scatter in the EFT parameter distributions, which translates into weaker improvements in cosmological parameter precision.

\begin{table*}[!t]
    \centering
    %\resizebox{\linewidth}{!}{
    \begin{tabular}{lcccccc}
    \toprule
    Dataset 
    & $w_0$ 
    & $w_a$ 
    & $\Omega_m$ 
    & $H_0$ 
    & $\sigma_8$  
    & $S_8$  \\
    \hline
    %\enspace
\textbf{SBP} &  &  &  &  &  & \\
$\cmb+\bao+P_\ell$ 
& $-0.81_{-0.15}^{+0.12}$ 
& $-0.52_{-0.31}^{+0.41}$ 
& $0.317_{-0.016}^{+0.012}$ 
& $66.76_{-1.39}^{+1.57}$ 
& $0.794_{-0.013}^{+0.013}$ 
& $0.8163_{-0.0109}^{+0.0100}$ %update
    \\
    $\cmb+\bao+P_\ell+\sn$ 
    & $\enspace -0.888_{-0.049}^{+0.049}\enspace$ 
& $\enspace -0.32_{-0.16}^{+0.16}\enspace$ 
& $\enspace 0.3095_{-0.0053}^{+0.0054}\enspace$ 
& $\enspace 67.52_{-0.55}^{+0.55}\enspace$ 
& $\enspace 0.8000_{-0.0075}^{+0.0074}\enspace$ 
& $\enspace 0.8125_{-0.0079}^{+0.0079}\enspace$ %final
    \\\hline
    \textbf{CBP} &  &  &  &  &  & \\
    $\cmb+\bao+P_\ell+B_0$ 
    & $-0.63_{-0.22}^{+0.16}$ 
& $-1.19_{-0.45}^{+0.62}$ 
& $0.329_{-0.022}^{+0.016}$ 
& $65.85_{-1.71}^{+2.00}$ 
& $0.796_{-0.014}^{+0.016}$ 
& $0.833_{-0.013}^{+0.012}$ %final
    \\
    $\cmb+\bao+P_\ell+B_0+\sn$
    & $-0.842_{-0.052}^{+0.052}$ 
& $-0.59_{-0.19}^{+0.19}$ 
& $0.3081_{-0.0053}^{+0.0053}$ 
& $67.84_{-0.55}^{+0.55}$ 
& $0.8105_{-0.0076}^{+0.0077}$ 
& $0.8214_{-0.0083}^{+0.0083}$ %final
    \\\hline    
    $\cmb+\bao$
    & $-0.43_{-0.21}^{+0.22}$ 
& $-1.71_{-0.63}^{+0.61}$ 
& $0.351_{-0.022}^{+0.022}$ 
& $63.86_{-2.10}^{+1.70}$ 
& $0.784_{-0.017}^{+0.015}$ 
& $0.847_{-0.012}^{+0.013}$
    \\\hline
    \end{tabular}
    %}
    \caption{\textbf{Dynamical dark energy}: Mean and 68\% confidence intervals on cosmological parameters in the $\wa$ analyses. The results are obtained using conservative Bayesian priors (CBP) and simulation-based priors (SBP). Constraints on the dark energy parameters and $\Omega_m$ are visualized in Figs.~\ref{fig:w0wa}\,\&\,\ref{fig:w0wa_Om} respectively.
    }
    \label{tab:w0wa}
\end{table*}

To understand the shifts more quantitatively, we utilize frequentist metrics following the DESI analyses~\cite{DESI:2025zgx}. Tab.~\ref{tab:chi2} lists the significance of the preference for the $\wa$ model over $\ld$, based on the $\Delta\chi^2_{\rm MAP}$, evaluated at the maximum a posteriori (MAP) points.
\begin{table}[!t]
\centering
%\resizebox{\columnwidth}{!}{
    \begin{tabular}{lccc}
    \toprule
    %& $\Delta$(DIC)
    Datasets & $\Delta\chi^2_{\rm MAP}$ & Significance  \\
    \hline
    % baseline
    \textbf{SBP} &  &  \\
    $\cmb+\bao+P_\ell$  
    & $-7.12$ 
    & $2.19\sigma$ \\
    $\cmb+\bao+P_\ell+\sn$  
    & $-3.80$ 
    & $1.44\sigma$ \\\hline
    % baseline
    \textbf{CBP} &  &  \\
    $\cmb+\bao+P_\ell+B_0$  
    & $-10.89$ 
    & $2.85\sigma$ \\
    $\cmb+\bao+P_\ell+B_0+\sn$ 
    & $-11.36$ 
    & $2.93\sigma$ \\
    \hline
    \end{tabular}
%}
\caption{\textbf{Preference for dynamical dark energy}: Difference in the $\chi^2_{\rm MAP}$ value for the best-fit $\wa$ model relative to the best-fit $\ld$ model with $w_0=-1$ and $w_a=0$ for fits to different combinations of datasets (as indicated). The third column reports the corresponding (frequentist) significance levels, expressed in terms of a $\sigma$-interval given the two additional free parameters. The results are shown for two choices of EFT parameter priors: conservative Bayesian priors (CBP) and simulation-based priors (SBP).
\label{tab:chi2}
}
\end{table}
With SBPs, the preference for the $\wa$ model in the $\cmb+\bao+P_\ell$ analysis is $2.2\sigma$, which is more than $0.6\sigma$ smaller than in the standard analysis using conservative EFT priors.
When supernova information is included, the evidence for evolving dark energy further decreases further to $1.4\sigma$ (with SBPs), which is $1.5\sigma$ lower than the result obtained with conservative priors. Finally, the addition of the galaxy bispectrum monopole to the SBP dataset lowers the preference for dynamical dark energy again to $1.2\sigma$, as shown in Appendix~\ref{app:bisp}.

These conclusions carry an important caveat. The results of the Bayesian SBP $\wa$ analyses that do not include supernova distance measurements are partially influenced by marginalization projection effects, which arise at the level of $0.5\sigma$ (consistent with \citep{DESI:2025wzd}, and much less than the $\approx 2\sigma$ obtained with na\"ive conservative priors without Alcock-Paczynski rescalings \citep{desi2,Tsedrik:2025hmj}).
As shown in Appendix~\ref{app:marg}, these ``prior volume'' effects pull the $w_0$ and $w_a$ posteriors towards the $\ld$ point ($w_0=-1$, $w_a=0$), sourcing a marked shift towards a cosmological constant.

A similar trend can be seen from comparing the Bayesian $w_0$-$w_a$ posteriors (Fig.~\ref{fig:w0wa}) to the $\chi^2_{\rm MAP}$ results (Tab.~\ref{tab:chi2}), which do not involve marginalization over nuisance parameters. For the $\cmb+\bao+P_\ell$ analysis, we find consistency with the cosmological constant at 68\% CL, but a $2.2\sigma$ deviation in the frequentist metric. 
This suggests that the marginalized posteriors in the $w_0$-$w_a$ plane are non-trivially impacted by projection effects arising from the marginalization over nuisance parameters~\cite{Chudaykin:2024wlw,Paradiso:2024yqh,Tsedrik:2025hmj}, in particular parameters such as $b_{\Gamma_3}$, which are not well constrained by the SBPs (see Appendix~\ref{app:params}).
When supernova information is included, we break degeneracies in the expansion history, which mitigates projection effects~\cite{DESI:2024hhd} ensuring that the SBP results are robust.

Our constraints imply that the evidence for evolving dark energy is not statistically significant, at least under the assumptions baked into our HOD priors. In addition, they show that the combination of full-shape datasets and SBPs can place tight constraints on time-varying dark energy scenarios, additionally implying that constraints from previous full-shape analyses adopting conservative EFT priors were substantially degraded by the exploration of EFT parameter regions that are inconsistent with HOD models. 

The above results can be contrasted with those of full-shape analyses using conservative EFT priors~\cite{DESI:2024hhd,desi2}. In this case, adding the baseline full-shape dataset to the BAO and CMB likelihood does not change the preference for evolving dark energy, due to the large parameter degeneracies found therein.\footnote{When incorporating additional datasets, such as the one-loop bispectrum, cross-correlations with CMB lensing, and the DESI photometric sample, full-shape information can have a non-negligible impact on cosmological constraints even when adopting conservative EFT priors~\citep{desi4}.}
Here we have demonstrated that incorporating small-scale information using SBPs allows one to unlock the full constraining power of full-shape clustering statistics in extended cosmological models, though the conclusions remain sensitive to modeling assumptions.

Finally, we discuss the constraints on $\Omega_m$, as illustrated in Fig.~\ref{fig:w0wa_Om}. Without relying on supernova information, adding full-shape information tightens the $\Omega_m$ posterior by $11\%$ when adopting conservative EFT priors and by $31\%$ with SBPs, due to the broken bias parameter degeneracies. Importantly, the central value of $\Omega_m$ shifts lower by $1.0\sigma$ and $1.6\sigma$ for CBPs and SBPs, respectively, bringing it into closer agreement with the DESI-only $\ld$ result~\cite{DESI:2025zgx}. When the SN dataset is included, SBPs do not improve the $\Omega_m$ constraint, implying that the expansion history in the $\wa$ model is already tightly constrained by supernova distance measurements~\cite{DESI:2024hhd,desi2}, as found for the $\ld$ model. 

\subsection{Neutrino masses}
\begin{figure*}
\includegraphics[width=0.99\columnwidth]{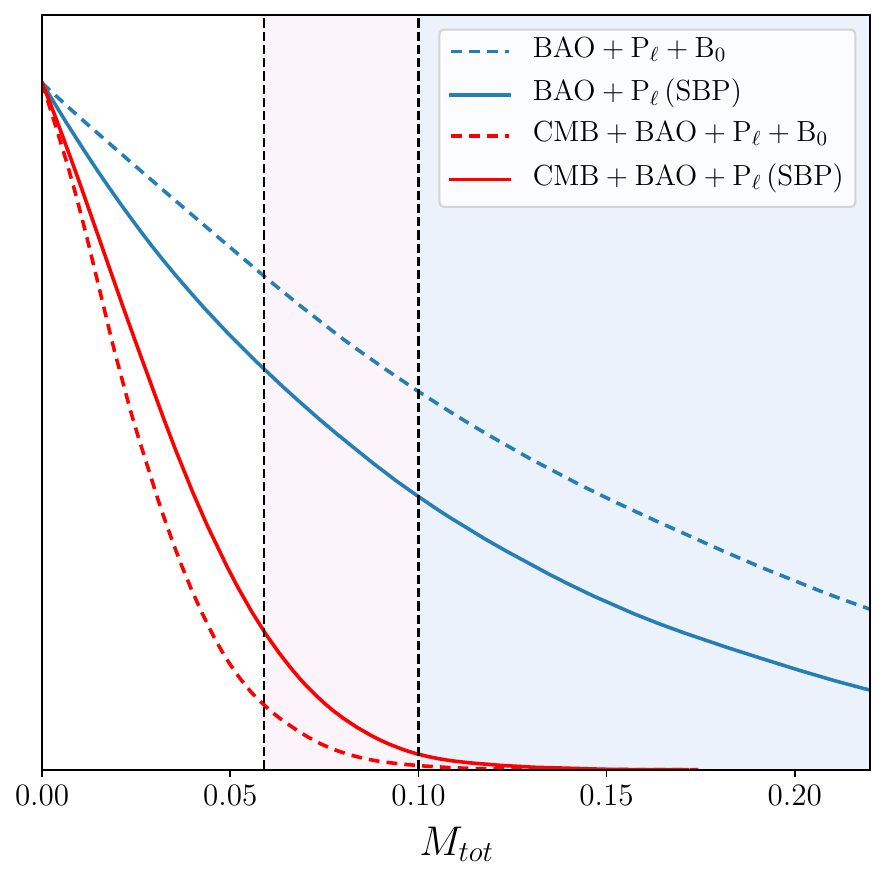}
\includegraphics[width=0.99\columnwidth]{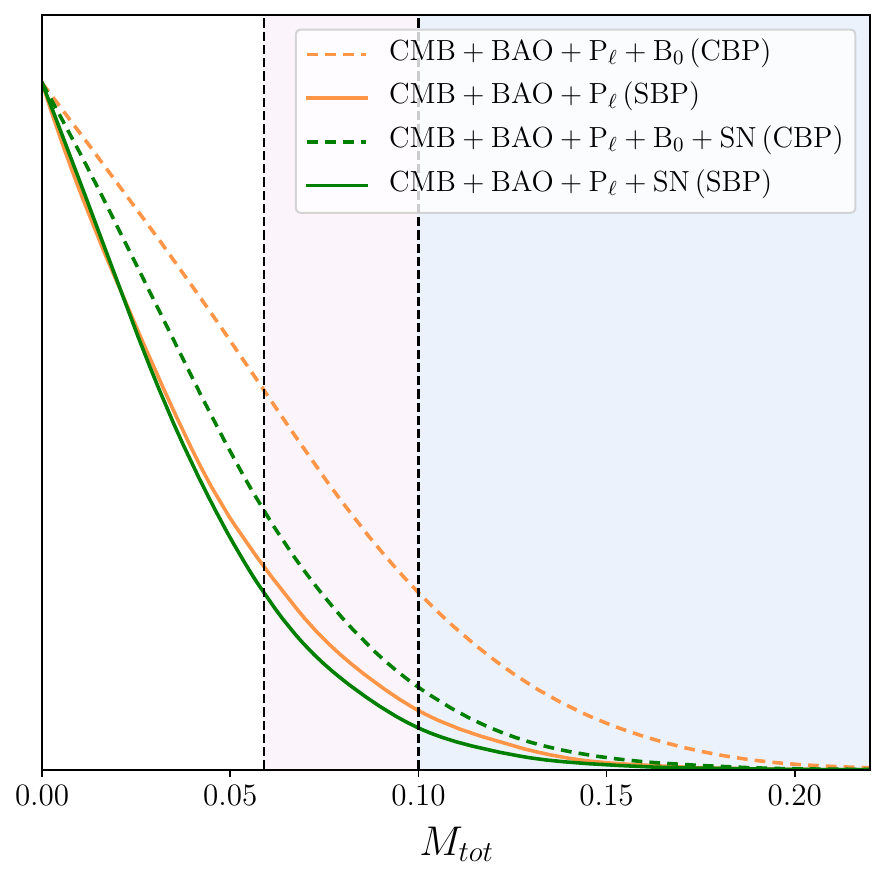}
\caption{\textbf{Massive neutrinos}: One-dimensional marginalized posterior distributions for the sum of neutrino masses when fixing the background model to $\ld$ ({\it left panel}) and $\wa$ ({\it right panel}). {\it Left panel:} $M_\nu+\ld$ analyses using the DESI DR2 BAO measurements, the DESI DR1 power spectrum $P_\ell$ and bispectrum $B_0$, and \textit{Planck} CMB data plus \textit{Planck} and ACT lensing (red). {\it Right panel:} Combined constraints on neutrino masses in the $M_\nu+\wa$ analyses using the conservative EFT priors (dashed) and SBPs (solid).}
\label{fig:mnu}
\end{figure*}

\noindent Lastly, we present constraints on the sum of neutrino masses, $M_\nu$. We explore parameter constraints in two different cosmological backgrounds, specified by the $\ld$ and $\wa$ models. 
As discussed in \paperfour, the latter constraints effectively marginalize over tensions between cosmological datasets that arise either due to background mismodeling or systematics in the data, see~\cite{Planck:2013nga,RoyChoudhury:2019hls,Craig:2024tky,Loverde:2024nfi,Green:2024xbb,Elbers:2024sha,Graham:2025fdt}
for detailed discussions. One-dimensional
marginalized constraints are given in Tab.~\ref{tab:mnu}, and visualized in Fig.~\ref{fig:mnu} for the $\ld$ (left panel) and $\wa$ (right panel) cosmological backgrounds.
\begin{table*}[!t]
    \centering
    %\resizebox{\linewidth}{!}{
    \begin{tabular}{lccccccc}
    \toprule
    Model/Dataset 
    & $M_\nu\,{\rm [eV]}$ 
    & $\Omega_m$ 
    & $H_0$ 
    & $\sigma_8$ 
    & $S_8$
    & $w_0$ 
    & $w_a$
    \\
    \hline
    %\enspace
    $\bm{\Lambda}$\textbf{CDM+}$\bm{M_\nu}$ &  &  &  &  &  & & \\
    %\textbf{SBP} &  &  &  &  &  & & \\
    $\bao+P_\ell$ (SBP)
    & $<0.262$ 
& $0.2992_{-0.0070}^{+0.0063}$ 
& $68.74_{-0.39}^{+0.39}$ 
& $0.765_{-0.016}^{+0.016}$ 
& $0.763_{-0.018}^{+0.018}$ %final
& $-$
& $-$
    \\
    $\cmb+\bao+P_\ell$ (SBP)
    & $<0.0730$ 
& $0.2983_{-0.0036}^{+0.0035}$ 
& $68.68_{-0.29}^{+0.29}$ 
& $0.8125_{-0.0054}^{+0.0065}$ 
& $0.8102_{-0.0075}^{+0.0076}$ %final
& $-$
& $-$
    \\\hline
    %\hline
    %\textbf{CBP} &  &  &  &  &  & & \\
    $\bao+P_\ell+B_0$ (CBP)
    & $<0.320$ 
& $0.2977_{-0.0070}^{+0.0070}$ 
& $68.71_{-0.61}^{+0.61}$ 
& $0.813_{-0.030}^{+0.030}$ 
& $0.810_{-0.031}^{+0.031}$ %final
& $-$
& $-$
    \\
    $\cmb+\bao+P_\ell+B_0$ (CBP)
    & $<0.0592$ 
& $0.2970_{-0.0035}^{+0.0035}$ 
& $68.79_{-0.29}^{+0.29}$ 
& $0.8167_{-0.0056}^{+0.0062}$ 
& $0.8126_{-0.0076}^{+0.0076}$ %final
& $-$
& $-$
    \\\hline
    $\bm{w_0w_a}$\textbf{CDM+$\bm{M_\nu}$} &  &  &  &  &  & & \\
    $\cmb+\bao+P_\ell+B_0$ (SBP)
    & $<0.100$ 
& $0.316_{-0.016}^{+0.012}$ 
& $66.84_{-1.37}^{+1.50}$ 
& $0.798_{-0.013}^{+0.013}$ 
& $0.818_{-0.010}^{+0.010}$ 
& $-0.82_{-0.15}^{+0.12}$ 
& $-0.45_{-0.31}^{+0.41}$  %final
    \\
    $\cmb\!+\!\bao\!+\!P_\ell\!+\!B_0\!+\!\sn$ (SBP)
    & $\enspace <0.0903\enspace$ 
& $\enspace 0.3086_{-0.0055}^{+0.0055}\enspace$ 
& $\enspace 67.56_{-0.57}^{+0.57}\enspace$ 
& $\enspace 0.8033_{-0.0081}^{+0.0081}\enspace$ 
& $\enspace 0.8147_{-0.0083}^{+0.0082}\enspace$ 
& $ -0.896_{-0.049}^{+0.049}$ 
& $ -0.26_{-0.16}^{+0.18}$ %final
    \\\hline
    $\cmb+\bao+P_\ell+B_0$ (CBP)
    & $<0.130$ 
& $0.327_{-0.023}^{+0.015}$ 
& $66.01_{-1.68}^{+2.05}$ 
& $0.799_{-0.015}^{+0.018}$ 
& $0.833_{-0.013}^{+0.012}$ 
& $-0.65_{-0.23}^{+0.16}$ 
& $-1.11_{-0.44}^{+0.66}$ %final
    \\
    $\cmb\!+\!\bao\!+\!P_\ell\!+\!B_0\!+\!\sn$ (CBP)
    & $<0.104$ 
& $0.3076_{-0.0054}^{+0.0054}$ 
& $67.88_{-0.56}^{+0.56}$ 
& $0.8136_{-0.0085}^{+0.0084}$ 
& $0.8237_{-0.0087}^{+0.0087}$ 
& $-0.850_{-0.052}^{+0.053}$ 
& $-0.54_{-0.18}^{+0.21}$ %final
    \\
    \hline
    \end{tabular}
    %}
    \caption{\textbf{Massive neutrinos}: Constraints on cosmological parameters when allowing the sum of neutrino masses to vary, assuming $M_\nu>0$ prior. We quote the mean and 68\% confidence intervals for all parameters, except $M_\nu$ for which we provide 95\% upper limits. The neutrino constraints are visualized in Fig.\,\ref{fig:mnu}.
    \label{tab:mnu}
    }
\end{table*}

In the $\Lambda$CDM model, 
our SBP-derived neutrino mass bounds are somewhat weaker than those 
from the conservative analyses. We attribute this to the lower $\sigma_8$ inference
in the SBP analyses, which allows for larger neutrino masses. In the $w_0w_a$CDM background, 
however, the situation is different: the SBP bounds on $M_\nu$ are substantially stronger
than the conservative ones. 
In particular, even in the absence of supernova data, we find $M_\nu<0.1$ eV at 95\% CL, coincident with the threshold separating the
two relevant neutrino mass hierarchies. When supernova data are included, we obtain $M_\nu<0.090$ eV,
which formally disfavors the
inverted hierarchy  
at around 2$\sigma$ level. 
This limit is similar to that of the ``kitchen sink" analysis of \paperfour, illustrating that SBPs carry significant additional constraining power.

\section{Conclusions}
\label{sec:concl}

\noindent In this work, we have presented a re-analysis of the public DESI DR1 full-shape galaxy clustering data using simulation-based priors on all EFT parameters entering the galaxy power spectrum. Our analysis shows that physically motivated priors derived from state-of-the-art halo occupation distribution models can significantly enhance 
cosmological information 
extracted from 
galaxy clustering data. 
Our approach based on 
normalizing flows~\cite{Sullivan:2021sof,Ivanov:2024hgq} 
allows one to accurately 
model intricate correlations between EFT parameters and capture small-scale clustering information that is typically marginalized over in analyses using conservative priors. 

Our SBPs are constructed from a broad class of HOD models including assembly bias (except for quasars), in order to minimize model-misspecification. When applied to our custom full-shape EFT likelihood, these priors lead to a significant improvement 
of cosmological constraints. Specifically, the errorbars on the clustering amplitude parameter $\sigma_8$ shrink by a factor of two, which is equivalent to increasing the effective survey volume by a factor of four, though this is associated with a sharp downwards shift in the posterior. In the standard $\Lambda$CDM model, we obtain percent-level constraints on $\Omega_m$, $H_0$, and $\sigma_8$ when combining full-shape data with DR2 BAO and the BBN prior. Our $\Lambda$CDM result for $\sigma_8=0.766\pm 0.015$ 
is in a mild $2.3\sigma$ tension with the ``kitchen sink'' analysis of \paperfour that reported $\sigma_8=0.815\pm 0.016$. 
\resub{This shift is mainly attributed to the quasar SBPs, which are based on more restrictive HOD models.}
For now, it is unclear whether this represents a breakdown of our EFT prior assumptions, systematics (as discussed below), or a statistical fluctuation; upcoming data from DESI DR2, \resub{together with quasar SBPs based on more general empirical models should shed light on this.} 

The impact of the SBPs 
extends beyond the standard cosmological model. 
In particular, in the $w_0w_a$CDM model, the priors improve the dark energy figure of merit by 70\% in combination with CMB data and by more than $20\%$ when further adding information from supernovae. Moreover, SBPs significantly reduce the apparent preference for dynamical dark energy to $2.2\sigma$ without supernova information and to $1.4\sigma$ when supernovae are included. In the latter case, the preference for dynamical dark energy is not statistically significant. 
Likewise, we find strong constraints on the sum of neutrino masses, producing the most stringent limit 
to date in the $w_0w_a$CDM background,
which is stronger
than the ``kitchen sink'' analysis 
of~\paperfour,
though a somewhat degraded constraint when assuming the $\ld$ background. Interestingly, the difference between neutrino mass limits in $\Lambda$CDM and $w_0w_a$CDM models reduce in the presence of SBPs, though it is difficult to interpret these shifts due to the one-sided neutrino mass posterior.

We note that our results are quite different from 
the official SBP collaboration analysis~\citep{DESI:2025wzd}.
In contrast to our work,~\citep{DESI:2025wzd} used more restrictive HOD models without any assembly bias, as well as informative Gaussian priors on EFT parameters. Like us, \citep{DESI:2025wzd} also used the normalizing flow modeling for SBPs~\cite{Sullivan:2021sof,Ivanov:2024hgq}, but found more modest 
improvements, with $\Omega_m$ and $\sigma_8$ contours tightening by $4\%$ and $23\%$, with virtually no improvement found for other parameters, including the dark energy equation of state. In contrast, our analysis found substantial improvements on $\sigma_8$, $H_0$, $w_0$ and $w_a$. We attribute this to the differences in the HOD parameter construction; \citep{DESI:2025wzd}
measured EFT parameters from galaxy power spectra extracted from simulations, whilst our work uses precise measurements from the field level, which do not inherit power spectrum degeneracies.
A similar conclusion was obtained in~\cite{Chen:2025jnr}, which showed than BOSS DR12 full-shape measurements enhanced with HOD-based priors can provide strong constraints on dynamical dark energy. 

It is important to reflect on the limitations of the HOD priors used in this work. As well as the intrinsic shortcomings of any HOD-based model (such as the simplified treatment of assembly bias and galaxy formation physics), these include the assumption of fixed $\ld$ cosmology in the HOD prior generation. While previous works have indicated that this is appropriate given the wide range of HOD parameters explored, high-fidelity tests of the priors in non-standard (and non-$\ld$) cosmologies have yet to be performed. Furthermore, we caution that our treatment does not account for any modulations of the EFT parameters caused by observational and systematic effects, such as fiber assignment, imaging weights, hydrodynamic feedback and beyond (though~\cite{Ivanov:2024dgv} found consistency between our HOD results and MilleniumTNG~\cite{Pakmor:2022yyn} and Astrid simulations~\cite{2024arXiv240910666N}). Usually, the EFT-based model is insensitive to such phenomena, since we marginalize over the free parameters with broad priors. In contrast, the SBPs build in physical relationships between EFT parameters that hold in systematic-free simulations but possibly not in contaminated data, implying that our overall constraints are less robust to systematics. We hope to address this issue more quantitatively in future.

Our work can be extended in multiple ways. One option is to explore alternative approaches to modeling the EFT parameter distributions along the lines of~\cite{Chen:2025jnr}. 
One can also extend the SBP analysis to the one-loop
bispectrum and two-dimensional clustering 
observables considered in \paperfour, which will further improve the parameter constraints.
Finally, one can re-analyze DESI full-shape 
data with SBPs in the context of 
non-minimal cosmological models such as those discussed in \citep{Ivanov:2020ril,DAmico:2020ods,Ibanez:2024uua,Ivanov:2021fbu,Ivanov:2021zmi,Chudaykin:2022nru,Ivanov:2021kcd,Ivanov:2023qzb,Cabass:2024wob,Chen:2022jzq,Cabass:2022ymb,Cabass:2022epm,Cabass:2024wob,Chen:2021wdi,Philcox:2021kcw,Wadekar:2020hax,Colas:2019ret,Ivanov:2019hqk,Rogers:2023ezo,Chen:2024vuf,He:2023oke,He:2023dbn,Xu:2021rwg,Chudaykin:2022nru,Chudaykin:2020ghx,McDonough:2023qcu,Toomey:2025yuy,Silva:2025twg,Chen:2024vuf,Maus:2025rvz}.

More broadly, our results illustrate how 
simulation-based information can be 
combined with EFT in actual data analyses in an economical way. Looking ahead to Stage-IV and Stage-V surveys, such as DESI-II, Euclid, Rubin, and Roman, one may argue that the role of the simulation based priors will only grow in future. As statistical uncertainties keep decreasing, cosmological inference will be more and more limited by theory systematics and modeling 
uncertainties. SBPs offer a flexible and scalable 
framework for encountering these challenging and thus fully realizing the scientific potential
of upcoming large-scale  
structure surveys.

\vskip 4pt
\textit{Acknowledgments.} 
{\small
\begingroup
\hypersetup{hidelinks}
We thank Adam Riess, Jamie Sullivan, and Mike Toomey for helpful
discussions. 
AC acknowledges funding from the Swiss National Science Foundation.
OHEP thanks the Kavli Institute for Theoretical Physics and its \href{https://www.flickr.com/photos/198816819@N07/55105931068}{pellagic collleagues}.
\endgroup

This research used data obtained with the Dark Energy Spectroscopic Instrument (DESI). DESI construction and operations is managed by the Lawrence Berkeley National Laboratory. This material is based upon work supported by the U.S. Department of Energy, Office of Science, Office of High-Energy Physics, under Contract No. DE–AC02–05CH11231, and by the National Energy Research Scientific Computing Center, a DOE Office of Science User Facility under the same contract. Additional support for DESI was provided by the U.S. National Science Foundation (NSF), Division of Astronomical Sciences under Contract No. AST-0950945 to the NSF’s National Optical-Infrared Astronomy Research Laboratory; the Science and Technology Facilities Council of the United Kingdom; the Gordon and Betty Moore Foundation; the Heising-Simons Foundation; the French Alternative Energies and Atomic Energy Commission (CEA); the National Council of Humanities, Science and Technology of Mexico (CONAHCYT); the Ministry of Science and Innovation of Spain (MICINN), and by the DESI Member Institutions: \url{www.desi.lbl.gov/collaborating-institutions}. The DESI collaboration is honored to be permitted to conduct scientific research on I’oligam Du’ag (Kitt Peak), a mountain with particular significance to the Tohono O’odham Nation. Any opinions, findings, and conclusions or recommendations expressed in this material are those of the author(s) and do not necessarily reflect the views of the U.S. National Science Foundation, the U.S. Department of Energy, or any of the listed funding agencies.
}

\section*{Data availability}\label{app:data0}
The data are currently not openly available but will be released upon completion of the set of related papers.

\appendix 

\section{EFT parameter constraints}\label{app:params}

\noindent In this Appendix, we present marginalized constraints on all EFT parameters in the $\ld$ analyses obtained using both CBPs and SBPs. In the first case, parameters that enter the model linearly are marginalized over analytically, thus their posteriors are derived from the chains {\it a posteriori}. 
In the SBP case, all EFT parameters were directly sampled in the MCMC chains. 
Results for the $P_\ell$ (CBP), $P_\ell+B_0$ (CBP) and $P_\ell$ (SBP) analyses are shown in Tab.~\ref{tab:EFT}. 

\begin{table*}[!t]
    \centering
    \footnotesize
  \begin{tabular}{cccc @{\hspace{1em}}|@{\hspace{1em}} cccc} \hline
    Param. 
    & $P_\ell$ 
    & $P_\ell+B_0$ 
    & $P_\ell$ (SBP)
    & Param.
    & $P_\ell$ 
    & $P_\ell+B_0$ 
    & $P_\ell$ (SBP)
    \\
    \hline
$\enspace b^{(1)}_1\enspace$
& $\enspace 1.60_{-0.10}^{+0.11}\enspace$
& $\enspace 1.729_{-0.081}^{+0.081}\enspace$
& $\enspace 1.849_{-0.058}^{+0.057}\enspace$
& $\enspace b^{(4)}_1\enspace$
& $\enspace 2.02_{-0.14}^{+0.16}\enspace$
& $\enspace 2.298_{-0.094}^{+0.094}\enspace$
& $\enspace 2.453_{-0.069}^{+0.068}\enspace$ \\
$b^{(1)}_2$
& $-0.83_{-1.62}^{+0.73}$
& $-0.71_{-0.74}^{+0.59}$
& $-0.19_{-0.32}^{+0.21}$
& $b^{(4)}_2$
& $-4.04_{-1.58}^{+1.16}$
& $0.43_{-1.28}^{+1.04}$
& $0.90_{-0.44}^{+0.63}$ \\
$b^{(1)}_{\mathcal{G}_2}$
& $-0.06_{-0.47}^{+0.43}$
& $-0.38_{-0.40}^{+0.40}$
& $-0.02_{-0.24}^{+0.21}$
& $b^{(4)}_{\mathcal{G}_2}$
& $-0.94_{-0.60}^{+0.71}$
& $-0.39_{-0.53}^{+0.53}$
& $-0.04_{-0.64}^{+0.35}$ \\
$b^{(1)}_{\Gamma_3}$ %bgs
& $-0.15_{-0.53}^{+0.53}$ %%
& $0.75_{-0.53}^{+0.53}$ %%
& $-0.35_{-1.03}^{+0.90}$
& $b^{(4)}_{\Gamma_3}$ %lrg3
& $0.60_{-0.74}^{+0.74}$ %%
& $0.54_{-0.73}^{+0.73}$ %%
& $-0.40_{-2.16}^{+1.78}$ \\
$c^{(1)}_{0}$ %bgs
& $-5.51_{-14.84}^{+14.84}$ %%
& $0.37_{-13.65}^{+13.65}$ %%
& $9.26_{-5.85}^{+5.13}$
& $c^{(4)}_{0}$ %lrg3
& $-4.20_{-15.54}^{+15.54}$ %%
& $-5.24_{-15.79}^{+15.79}$ %%
& $8.92_{-10.80}^{+11.13}$ \\
$c^{(1)}_{2}$ %bgs
& $31.17_{-12.95}^{+12.95}$ %%
& $47.68_{-12.95}^{+12.95}$ %%
& $58.31_{-13.09}^{+13.17}$
& $c^{(4)}_{2}$ %lrg3
& $33.59_{-12.80}^{+12.80}$ %%
& $70.92_{-12.62}^{+12.62}$ %%
& $66.12_{-17.91}^{+22.22}$ \\
$c^{(1)}_{4}$ %bgs
& $0.82_{-18.30}^{+18.30}$ %%
& $1.38_{-17.86}^{+17.86}$ %%
& $-71.66_{-10.02}^{+8.19}$
& $c^{(4)}_{4}$ %lrg3
& $30.22_{-18.93}^{+18.93}$ %%
& $37.71_{-19.40}^{+19.40}$ %%
& $-87.87_{-17.80}^{+12.38}$ \\
$b^{(1)}_4$ %bgs
& $403.75_{-335.21}^{+335.21}$ %%
& $423.85_{-303.57}^{+303.57}$ %%
& $589.26_{-263.05}^{+262.39}$
& $b^{(4)}_4$ %lrg3
& $67.92_{-212.22}^{+212.22}$ %%
& $72.64_{-164.21}^{+164.21}$ %%
& $789.46_{-98.02}^{+112.54}$ \\
$P^{(1)}_{\rm shot}$ %bgs
& $-0.15_{-0.38}^{+0.38}$
& $-0.22_{-0.35}^{+0.35}$
& $-0.34_{-0.23}^{+0.17}$
& $P^{(4)}_{\rm shot}$ %lrg3
& $-0.68_{-0.11}^{+0.11}$
& $-0.30_{-0.10}^{+0.10}$
& $-0.284_{-0.189}^{+0.089}$ \\
$a^{(1)}_0$ %bgs
& $-0.57_{-0.67}^{+0.67}$
& $-0.42_{-0.70}^{+0.70}$
& $-0.017_{-0.081}^{+0.076}$
& $a^{(4)}_0$ %lrg3
& $0.61_{-0.25}^{+0.25}$
& $-0.23_{-0.28}^{+0.28}$
& $-0.12_{-0.14}^{+0.14}$ \\
$a^{(1)}_2$ %bgs
& $0.08_{-0.76}^{+0.76}$
& $-0.00_{-0.80}^{+0.80}$
& $0.92_{-0.59}^{+0.73}$
& $a^{(4)}_2$ %lrg3
& $0.62_{-0.50}^{+0.50}$
& $0.38_{-0.56}^{+0.56}$
& $2.23_{-0.54}^{+0.53}$ \\\hline
$b^{(2)}_1$
& $1.79_{-0.11}^{+0.12}$
& $1.932_{-0.085}^{+0.078}$
& $2.047_{-0.058}^{+0.058}$
& $b^{(5)}_1$
& $1.26_{-0.11}^{+0.15}$
& $1.421_{-0.079}^{+0.079}$
& $1.558_{-0.047}^{+0.047}$ \\
$b^{(2)}_2$
& $-1.11_{-1.33}^{+0.86}$
& $-0.60_{-0.72}^{+0.62}$
& $-0.02_{-0.35}^{+0.25}$
& $b^{(5)}_2$
& $-0.85_{-5.63}^{+7.93}$
& $-1.42_{-1.44}^{+1.26}$
& $-0.68_{-0.20}^{+0.22}$ \\
$b^{(2)}_{\mathcal{G}_2}$
& $-0.31_{-0.44}^{+0.44}$
& $-0.41_{-0.40}^{+0.40}$
& $-0.13_{-0.24}^{+0.20}$
& $b^{(5)}_{\mathcal{G}_2}$
& $-0.91_{-0.99}^{+1.45}$
& $-0.60_{-0.67}^{+0.68}$
& $-0.14_{-0.28}^{+0.17}$ \\
$b^{(2)}_{\Gamma_3}$ %lrg1
& $0.52_{-0.58}^{+0.58}$ %%
& $0.67_{-0.57}^{+0.57}$ %%
& $0.08_{-0.96}^{+0.96}$
& $b^{(5)}_{\Gamma_3}$ %elg2
& $0.35_{-0.83}^{+0.83}$ %%
& $0.24_{-0.82}^{+0.82}$ %%
& $0.34_{-1.01}^{+1.63}$ \\
$c^{(2)}_{0}$ %lrg1
& $-9.51_{-14.33}^{+14.33}$ %%
& $-13.27_{-13.56}^{+13.56}$ %%
& $5.22_{-5.18}^{+5.14}$
& $c^{(5)}_{0}$ %elg2
& $-0.27_{-11.64}^{+11.64}$ %%
& $-5.94_{-11.77}^{+11.77}$ %%
& $0.60_{-2.01}^{+2.98}$ \\
$c^{(2)}_{2}$ %lrg1
& $44.99_{-12.13}^{+12.13}$ %%
& $53.94_{-12.14}^{+12.14}$ %%
& $39.05_{-12.73}^{+12.74}$
& $c^{(5)}_{2}$ %elg2
& $24.22_{-10.59}^{+10.59}$ %%
& $25.90_{-10.53}^{+10.53}$ %%
& $6.63_{-11.39}^{+9.12}$ \\
$c^{(2)}_{4}$ %lrg1
& $3.88_{-17.24}^{+17.24}$ %%
& $2.06_{-17.32}^{+17.32}$ %%
& $-87.51_{-9.54}^{+6.90}$
& $c^{(5)}_{4}$ %elg2
& $-14.00_{-15.15}^{+15.15}$ %%
& $-14.03_{-16.36}^{+16.36}$ %%
& $-12.01_{-6.23}^{+8.54}$ \\
$b^{(2)}_4$ %lrg1
& $324.08_{-269.66}^{+269.66}$ %%
& $329.46_{-226.53}^{+226.53}$ %%
& $823.32_{-149.59}^{+149.70}$
& $b^{(5)}_4$ %elg2
& $262.84_{-254.01}^{+254.01}$ %%
& $235.41_{-209.05}^{+209.05}$ %%
& $42.57_{-65.28}^{+47.06}$ \\
$P^{(2)}_{\rm shot}$ %lrg1
& $-0.75_{-0.32}^{+0.32}$
& $-0.85_{-0.30}^{+0.30}$
& $-0.41_{-0.23}^{+0.17}$
& $P^{(5)}_{\rm shot}$ %elg2
& $-0.405_{-0.042}^{+0.042}$
& $-0.140_{-0.039}^{+0.039}$
& $-0.036_{-0.035}^{+0.036}$ \\
$a^{(2)}_0$ %lrg1
& $0.05_{-0.60}^{+0.60}$
& $-0.17_{-0.64}^{+0.64}$
& $-0.036_{-0.081}^{+0.084}$
& $a^{(5)}_0$ %elg2
& $0.40_{-0.11}^{+0.11}$
& $-0.07_{-0.14}^{+0.14}$
& $0.011_{-0.018}^{+0.018}$ \\
$a^{(2)}_2$ %lrg1
& $-0.01_{-0.77}^{+0.77}$
& $-0.15_{-0.81}^{+0.81}$
& $0.99_{-0.61}^{+0.61}$
& $a^{(5)}_2$ %elg2
& $0.40_{-0.29}^{+0.29}$
& $0.51_{-0.35}^{+0.35}$
& $0.04_{-0.12}^{+0.11}$ \\\hline
$b^{(3)}_1$
& $1.93_{-0.12}^{+0.13}$
& $2.138_{-0.086}^{+0.085}$
& $2.261_{-0.061}^{+0.061}$
& $b^{(6)}_1$
& $2.06_{-0.16}^{+0.20}$
& $2.30_{-0.12}^{+0.12}$
& $2.522_{-0.075}^{+0.074}$ \\
$b^{(3)}_2$
& $-2.24_{-1.34}^{+0.92}$
& $0.63_{-0.94}^{+0.78}$
& $0.10_{-0.43}^{+0.26}$
& $b^{(6)}_2$
& $-5.21_{-3.71}^{+1.89}$
& $-2.69_{-3.21}^{+2.15}$
& $1.25_{-0.14}^{+0.14}$ \\
$b^{(3)}_{\mathcal{G}_2}$
& $-0.56_{-0.48}^{+0.53}$
& $0.18_{-0.44}^{+0.44}$
& $-0.36_{-0.30}^{+0.19}$
& $b^{(6)}_{\mathcal{G}_2}$
& $-1.74_{-1.19}^{+1.37}$
& $-1.20_{-0.85}^{+0.96}$
& $-0.884_{-0.100}^{+0.099}$ \\
$b^{(3)}_{\Gamma_3}$ %lrg2
& $0.20_{-0.64}^{+0.64}$ %%
& $-0.23_{-0.63}^{+0.63}$ %%
& $1.06_{-1.00}^{+1.35}$
& $b^{(6)}_{\Gamma_3}$ %qso
& $0.61_{-0.86}^{+0.86}$ %%
& $0.69_{-0.84}^{+0.84}$ %%
& $3.00_{-0.36}^{+0.36}$ \\
$c^{(3)}_{0}$ %lrg2
& $-5.05_{-14.90}^{+14.90}$ %%
& $-8.19_{-14.44}^{+14.44}$ %%
& $7.83_{-5.94}^{+5.96}$
& $c^{(6)}_{0}$ %qso
& $21.43_{-17.32}^{+17.32}$ %%
& $1.58_{-17.23}^{+17.23}$ %%
& $-0.02_{-3.97}^{+3.25}$ \\
$c^{(3)}_{2}$ %lrg2
& $32.12_{-12.00}^{+12.00}$ %%
& $47.74_{-11.93}^{+11.93}$ %%
& $37.30_{-15.91}^{+13.30}$
& $c^{(6)}_{2}$ %qso
& $36.09_{-14.71}^{+14.71}$ %%
& $39.65_{-14.80}^{+14.80}$ %%
& $18.22_{-11.10}^{+9.86}$ \\
$c^{(3)}_{4}$ %lrg2
& $-2.67_{-17.53}^{+17.53}$ %%
& $-5.01_{-17.51}^{+17.51}$ %%
& $-98.54_{-10.33}^{+6.60}$
& $c^{(6)}_{4}$ %qso
& $-11.51_{-21.43}^{+21.43}$ %%
& $-8.49_{-22.46}^{+22.46}$ %%
& $-34.01_{-9.91}^{+8.89}$ \\
$b^{(3)}_4$ %lrg2
& $320.62_{-218.24}^{+218.24}$ %%
& $345.63_{-175.77}^{+175.77}$ %%
& $844.53_{-114.15}^{+114.48}$
& $b^{(6)}_4$ %qso
& $65.08_{-221.28}^{+221.28}$ %%
& $40.18_{-176.12}^{+176.12}$ %%
& $201.63_{-75.21}^{+75.57}$ \\
$P^{(3)}_{\rm shot}$ %lrg2
& $-0.61_{-0.25}^{+0.25}$
& $-0.40_{-0.24}^{+0.24}$
& $-0.23_{-0.25}^{+0.24}$
& $P^{(6)}_{\rm shot}$ %qso
& $-0.139_{-0.011}^{+0.011}$
& $-0.058_{-0.011}^{+0.011}$
& $-0.0367_{-0.0048}^{+0.0048}$ \\
$a^{(3)}_0$ %lrg2
& $0.45_{-0.51}^{+0.51}$
& $-0.18_{-0.55}^{+0.55}$
& $-0.07_{-0.12}^{+0.13}$
& $a^{(6)}_0$ %qso
& $0.208_{-0.038}^{+0.038}$
& $0.081_{-0.044}^{+0.044}$
& $0.0609_{-0.0102}^{+0.0094}$ \\
$a^{(3)}_2$ %lrg2
& $-0.28_{-0.74}^{+0.74}$
& $-0.50_{-0.79}^{+0.79}$
& $1.04_{-0.64}^{+0.64}$
& $a^{(6)}_2$ %qso
& $-0.15_{-0.10}^{+0.10}$
& $-0.16_{-0.12}^{+0.12}$
& $-0.277_{-0.049}^{+0.044}$ \\
  \hline
    \end{tabular}
    \caption{\textbf{EFT parameter constraints}: EFT parameter constraints derived from the $\Lambda$CDM analyses of $P_\ell$ and $P_\ell+B_0$ using the conservative Bayesian priors (CBP) from \papertwo, and from $P_\ell$ when adopting the simulation-based priors of this work (SBP). In all cases, we present the mean values and 68\% confidence intervals. For CBPs, nuisance parameters that enter quadratically in the likelihood are marginalized over analytically, with their posteriors recovered from the chains {\it a posteriori}.
    The use of SBPs in most cases significantly sharpens constraints on the EFT parameters compared to the standard $P_\ell+B_0$ analysis employing CBPs. 
    \resub{The leading-order counterterms, $c_0$, $c_2$ and $c_4$, are reported in units of $[\Mpc/h]^2$, while the next-to-leading order redshift-space counterterm $b_4$ is expressed in units of $[\Mpc/h]^4$.}
    Superscripts
    (1)-(6) correspond to BGS, LRG1, LRG2, LRG3, ELG2, and QSO samples, respectively.
    }
\label{tab:EFT}
\end{table*}

\begin{table*}[ht]
    \centering
    %\resizebox{\linewidth}{!}{
    \begin{tabular}{lcccccccc}
    \toprule
    \textbf{Model/Mock data} 
    & $\omega_{\rm cdm}$
    & $H_0$ 
    & ${\ln(10^{10}A_s)}$ 
    & $n_s$
    & $\Omega_m$ 
    & $\sigma_8$  
    & $w_0$
    & $w_a$ \\
    \hline
    $\bm{\Lambda}$\textbf{CDM} &  &  &  &  &  & &  & \\
    $P_\ell$ (SBP)
    & $\enspace 0.01\sigma\enspace$
    & $\enspace -0.24\sigma\enspace$ 
    & $\enspace 0.24\sigma\enspace$ 
    & $\enspace 0.26\sigma\enspace$ 
    & $\enspace 0.20\sigma\enspace$ 
    & $\enspace 0.45\sigma\enspace$ 
    & $-$
    & $-$ % final
    \\
    $P_\ell+\bao$ (SBP)
    & $0.01\sigma$
    & $-0.04\sigma$ 
    & $0.24\sigma$ 
    & $0.35\sigma$ 
    & $0.02\sigma$ 
    & $0.44\sigma$ 
    & $-$
    & $-$ % final
    \\
    \hline
    $\bm{w_0w_a}$\textbf{CDM} &  &  &  &  &  & &  & \\
    $\cmb+P_\ell+\bao$ (SBP)
    & $-0.12\sigma$
    & $0.43\sigma$ 
    & $0.24\sigma$ 
    & $0.14\sigma$ 
    & $-0.40\sigma$ 
    & $0.37\sigma$ 
    & $\enspace -0.49\sigma\enspace$
    & $\enspace 0.50\sigma\enspace$ % update!
    \\
    \hline
    \end{tabular}
    %}
    \caption{\textbf{Projection effects}: Differences between the true parameter values and the posterior means recovered from analyses of synthetic mock data in the $\ld$ and $\wa$ models. We find good recovery of the input cosmological parameters indicating that prior volume effects are fairly well controlled.
    }
    \label{tab:marg}
\end{table*}

We begin by presenting the results obtained using conservative EFT priors. 
In agreement with \paperone, the inclusion of the bispectrum significantly tightens constraints on the linear and quadratic bias parameters, since these parameters enter the bispectrum theoretical model at leading-order.

SBPs significantly improve the constraints on most EFT parameters. This can be quantified by looking at the information gain on linear and quadratic bias parameters.
The $b_1$ posteriors shrink on average by $30\%$ for the BGS and LRG galaxy samples, and by $40\%$ for ELG2 and QSO tracers, relative to the standard $P_\ell+B_0$ analysis adopting conservative EFT priors.
The improvement for $b_2$ is more pronounced, with uncertainties reduced by factors of $2.1$ to $2.5$ for the galaxy samples, and by factors of $6$ and $21$ for the ELG2 and QSO samples, respectively.
The $b_{\mathcal{G}_2}$ error-bars also decrease substantially, by $40\%$ for the galaxy samples except LRG3, where the improvement is more modest, and by factors of $2.6$ and $9.2$ for the ELG2 and QSO tracers, respectively. 
A comparable level of improvement is observed for several other EFT parameters, including EFT counterterms and stochastic parameters.

The larger information gain for QSO nuisance parameters is partially attributed to the more restrictive HOD models (without assembly bias) used to generate the SBPs for this tracer, as discussed in Sec.~\ref{sec:hod}. 
However, we note that the QSO 
(and also ELG2) samples of the DESI DR1 data have lower signal-to-noise and thus only marginally contribute to the cosmological parameter constraints; see~\cite{desi1} for details, and as such, do not expect these samples to significantly affect our overall conclusions.

Interestingly, the use of SBPs considerably weakens the $b_{\Gamma_3}$ constraints relative to the standard $P_\ell+B_0$ analysis (for all tracers except QSO). This occurs since our SBPs are constructed using decorated HOD models that cannot constrain this parameter~\cite{Ivanov:2024hgq,Ivanov:2024xgb},
such that the conservative EFT priors yield a tighter constraint on the cubic bias parameter than the SBPs. 
In principle, $b_{\Gamma_3}$ can be better constrained by incorporating higher-order statistics, with the bispectrum breaking degeneracies between $b_{\mathcal{G}_2}$ and $b_{\Gamma_3}$. 
However, the DESI DR1 data provides limited signal-to-noise of the higher-order statistics on large scales~\cite{desi1}, so including the galaxy bispectrum improves the $b_{\Gamma_3}$ constraint only marginally when SBPs are employed: by $20
\%$ for the LRG3 sample and by $\simeq 2-7\,\%$ for the other DESI tracers. 
The situation may change with future DESI data releases, which are expected to increase the utility of the galaxy bispectrum within the SBP analysis. 
This will be explored in future work.

\section{Parameter projection effects}
\label{app:marg}

\begin{figure*}[!t]
	\centering
	\includegraphics[width=0.75\textwidth]{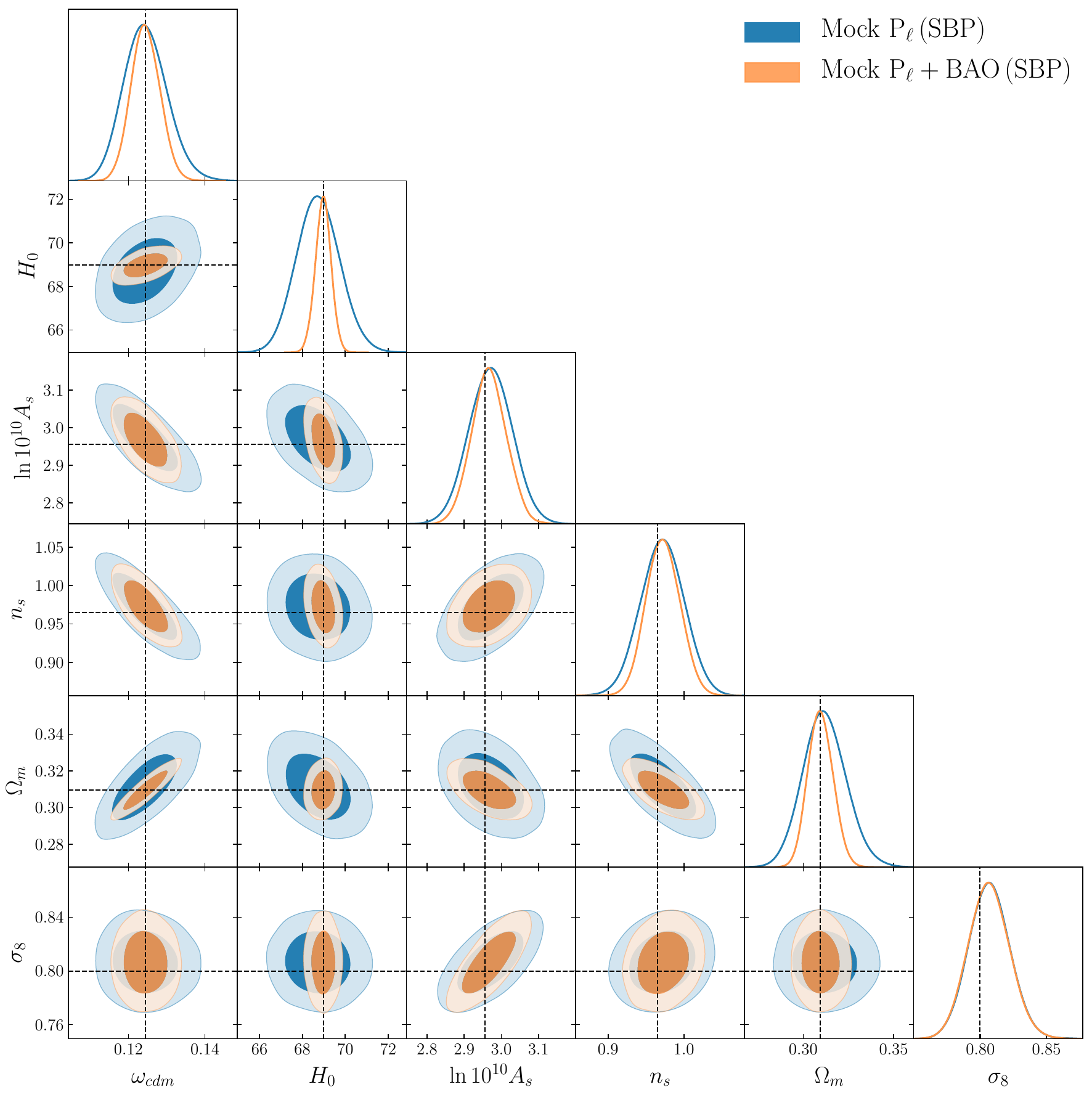}
	\caption{\textbf{Projection Effects in $\bm{\Lambda}$\textbf{CDM}}: Cosmological parameter constraints obtained from the synthetic mock data generated with our pipeline in the $\ld$ model using simulation-based priors (SBP). Results are shown for two datasets: $P_\ell$ (blue) and $P_\ell+\bao$ (orange). We do not find significant prior volume shifts in this case, except for a $0.45\sigma$ shift for $\sigma_8$.
     }
    \label{fig:synth_ld}
\end{figure*}

\begin{figure*}[!t]
	\centering
	\includegraphics[width=0.7\textwidth]{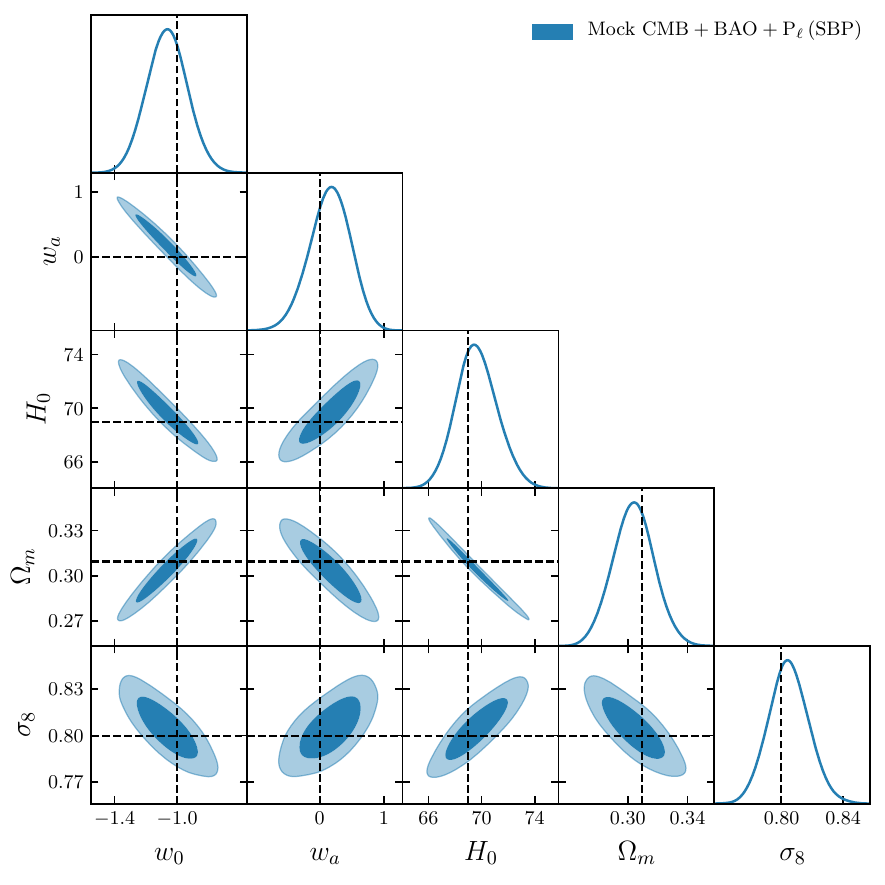}
	\caption{\textbf{Projection effects in $\bm{w_0w_a}$\textbf{CDM}}: As Fig.~\ref{fig:synth_ld} but for the $\wa$ model. 
    }
    \label{fig:synth_wa}
\end{figure*}

\noindent In this Appendix, we quantify the magnitude of parameter projection effects when using SBPs in the $\ld$ and $\wa$ models. Tab.~\ref{tab:marg} shows the impact of prior volume effects, with the one- and two-dimensional marginalized posterior distributions shown in Figs.~\ref{fig:synth_ld}\,\&\,\ref{fig:synth_wa}.

Following \paperone and \papertwo, we assess marginalization-induced biases in parameter recovery by generating a synthetic data vector with our theory pipeline. As a first step, we compute the best-fit SBP $\ld$ model with ($w_0=-1$, $w_a=0$) from the $\bao+P_\ell$ data by fitting the three cosmological parameters $\{\omega_{cdm}, H_0, \ln(10^{10}A_s)\}$, while fixing  $\omega_{b}=0.02218$, $n_s=0.9649$, $\tau=0.0544$, together with six copies of the parameter set
\begin{align}
    &\{b_1\sigma_8,b_2\sigma_8^2,b_{\G}\sigma_8^2,b_{\Gamma_3}A_{\rm amp}^2,c_0A_{\rm amp},c_2A_{\rm amp},c_4A_{\rm amp},\nonumber\\
    &\,\,\,b_4A_{\rm amp},P_{\rm shot},a_{0},a_{2}\},
\end{align}
corresponding to each DESI data chunk in the full-shape analysis. The EFT parameters are rescaled by the late-time fluctuation amplitude, $\sigma_{8}(z)$, via $A_{\rm amp}\equiv\sigma_8^2(z)/\sigma_{8,{\rm fid}}^2(z)$ where ``fid'' refers to quantities evaluated in the fiducial \textit{Planck} 2018 cosmology.
The resulting best-fit model is then used to generate mock galaxy power spectrum data at six different redshifts, simulating the DESI samples, as well as mock DESI BAO DR2 and CMB measurements. To construct the CMB mock data, we use the likelihood \textsc{fake\_planck\_realistic} implemented in the \textsc{Montepython} code~\cite{Audren:2012wb,Brinckmann:2018cvx}. In particular, we generate mock temperature, polarization and CMB lensing potential data, together with the corresponding noise spectra, adopting the recommended settings that reproduce the characteristics of the full Planck mission~\cite{Aghanim:2018eyx}. 

We fit the synthetic data using the same theoretical pipeline employed in the real data analysis. We adopt the same covariance matrices as in the data analyses, while for the CMB mock data the uncertainties are estimated from the simulated temperature and polarization noise spectra~\cite{Brinckmann:2018owf}. In analyses that exclude the CMB information, we fix $\omega_b$ and $\tau$ to the value used to generate the mock data and impose a wide Gaussian prior on $n_s$, replicating the setup of the real-data SBP analysis (see Sec.~\ref{sec:data2}). When the CMB data are included, we vary all cosmological parameters, following the main analyses of this work. 

We begin with the $\ld$ model.

In the $P_\ell$ analysis, the marginalization shift remains below $\approx0.25\sigma$ for all parameters, except $\sigma_8$, which demonstrates a larger marginalization bias at the level of $0.45\sigma$ (though in the opposite direction to the shift between SBPs and CBPs found in our main analyses). The addition of the mock BAO dataset significantly reduces parameter shifts for $H_0$ and $\Omega_m$, while the $\sigma_8$ constraint remains largely unaffected (as expected when adding distance information).
All in all, our analysis pipeline with SBPs produces fairly robust parameter constraints in the $\ld$ model. Notably, we do not combine our analysis with the galaxy bispectrum, as this does not bring much information when analyzed with SBPs, as shown in Appendix~\ref{app:bisp}.~\footnote{In contrast to BOSS data~\cite{Ivanov:2024dgv}, our parameter constraints remain largely unaffected when adding 
the galaxy bispectrum in the SBP analysis. We attribute this to differences in bias and number density between DESI and BOSS samples. }
We also do not consider CMB mock data, because the parameter projection effects are strongly suppressed in the $\ld$ model once CMB information is included~\cite{DESI:2024hhd}.

Next, we assess parameter projection effects in the $\wa$ scenario. In this case, we always combine the simulated full-shape statistics with mock BAO and CMB datasets, following the main analyses of this work. 
From Tab.~\ref{tab:marg} it is clear that all shifts are appreciably smaller than the statistical error-bars, with a maximal shift is about $0.5\sigma$ at the level of 
the $w_0$, $w_a$ parameters. This is similar to the marginalization shifts when using the conservative
priors rescaled with the Alcock-Paczynski parameters, as discussed in \papertwo (cf.\,\citep{DESI:2025wzd}).
We do not combine our mock likelihood with any SN data, since the parameter projection effects are strongly suppressed in the $\wa$ model once supernova distance information is included~\cite{DESI:2024hhd}.

It is interesting to compare our conclusions regarding projection effects to those obtained using the conservative EFT priors from \papertwo. For the $\ld$ model, we find that adopting SBPs significantly reduces the marginalization-induced shifts in the DESI-only analyses. In particular, in the $P_\ell$ analysis, the marginalization-induced shifts decreases from $0.37\sigma$ to $0.01\sigma$ for $w_{\rm cdm}$ and from $0.48\sigma$ to $0.2\sigma$ for $\Omega_m$, while the shift in the $\sigma_8$ recovery slightly increases from $0.15\sigma$ to $0.45\sigma$ when using SBPs~\cite{desi2}. 
In the $\wa$ model, in the $\cmb+\bao+P_\ell$ analysis, parameter projection effects remain largely unchanged when adopting SBPs instead of CBPs.
In particular, the marginalization-induced shift for $w_0$ slightly increases from $0.41\sigma$ to $0.49\sigma$, and from $0.37\sigma$ to $0.5\sigma$ for $w_a$, while the biases in the recovery of $\omega_{cdm}$, $H_0$, $\Omega_m$ and $\sigma_8$ decrease marginally relative to the conservative analysis~\cite{desi2}. 

\begin{table}[!t]
\centering
%\resizebox{\columnwidth}{!}{
    \begin{tabular}{lccc}
    \toprule
    Datasets & $\Delta\chi^2_{\rm MAP}$ & Significance  \\
    \hline
    \textbf{SBP} &  &  \\
    % w bispectrum
    $\cmb+\bao+P_\ell+B_0$  
    & $-8.91$ 
    & $2.52\sigma$ \\
    $\cmb+\bao+P_\ell+B_0+\sn$  
    & $-3.08$ 
    & $1.24\sigma$ \\\hline
    \end{tabular}
%}
\caption{\textbf{Preference for dynamical dark energy when including bispectra}: Difference in the $\chi^2_{\rm MAP}$ value for the best-fit $\wa$ model relative to the best-fit $\ld$ model with $w_0=-1$ and $w_a=0$ for fits to different combinations of datasets (as indicated). The third column reports the corresponding (frequentist) significance levels, expressed in terms of a $\sigma$-interval given the additional two free parameters. In all cases, we adopt simulation-based priors (SBP).
\label{tab:chi22}
}
\end{table}
\begin{table*}[!t]
    \centering
    %\resizebox{\linewidth}{!}{
    \begin{tabular}{lccccc}
    \toprule
    Dataset 
    & $w_0$ 
    & $w_a$ 
    & $\Omega_m$ 
    & $H_0$ 
    & $\sigma_8$  \\
    \hline
    %\enspace
    \textbf{GSBP} &  &  &  &  & \\
    $\cmb+\bao+P_\ell+B_0$ 
    & $\enspace -0.73_{-0.21}^{+0.16}\enspace$ 
& $\enspace -0.82_{-0.41}^{+0.57}\enspace$ 
& $\enspace 0.324_{-0.021}^{+0.016}\enspace$ 
& $\enspace 66.19_{-1.77}^{+1.95}\enspace$ 
& $\enspace 0.792_{-0.015}^{+0.015}\enspace$ %temp
    \\
    $\cmb+\bao+P_\ell+B_0+\sn$ 
    & $-0.875_{-0.053}^{+0.049}$ 
& $-0.42_{-0.17}^{+0.19}$ 
& $0.3094_{-0.0054}^{+0.0055}$ 
& $67.64_{-0.58}^{+0.57}$ 
& $0.8024_{-0.0078}^{+0.0077}$ %temp
    \\
    \hline
    \end{tabular}
    %}
    \caption{\textbf{Dynamical dark energy with Gaussian SBPs}: Mean and 68\% confidence intervals on cosmological parameters for the $\wa$ analyses, when using approximate Gaussian simulation-based priors (GSBP). 
    }
    \label{tab:w0wa2}
\end{table*}

\section{SBP analysis including the bispectrum monopole}
\label{app:bisp}

\begin{figure}
\includegraphics[width=0.99\columnwidth]{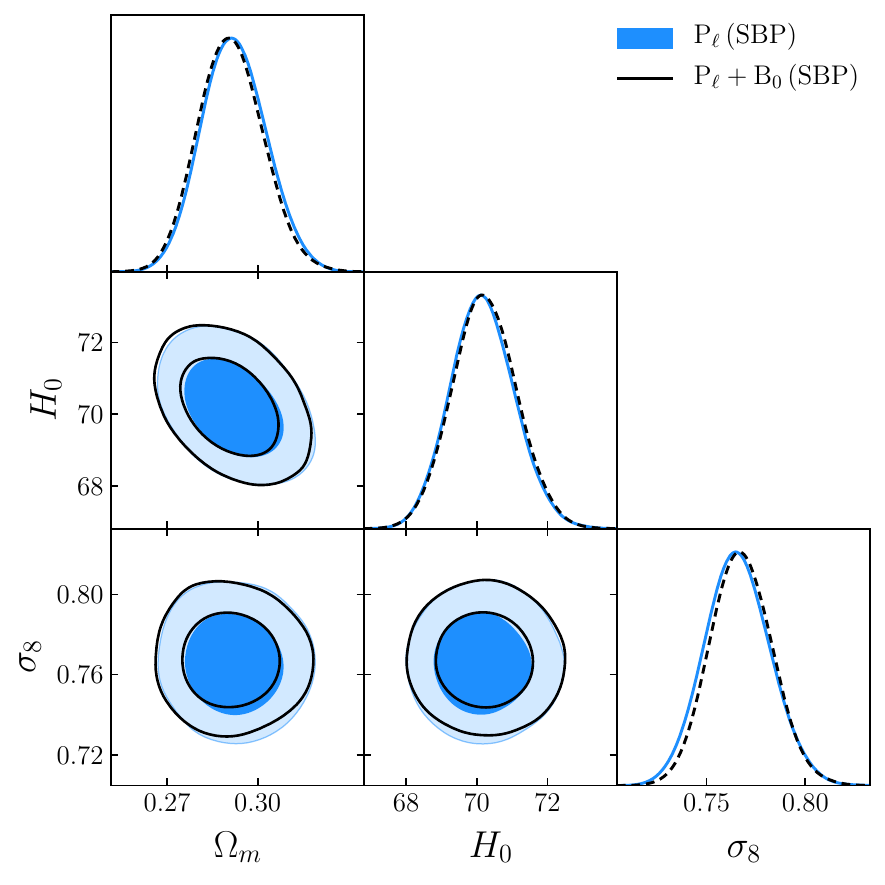}
\caption{\textbf{SBP analysis with galaxy bispectrum}: Posterior distribution of $\Omega_m$, $H_0$ and $\sigma_8$ obtained from the DESI DR1 power spectrum ($P_\ell$) alone and in combination with the DESI DR1 bispectrum monopole ($B_0$), both analyzed using simulation-based priors (SBP). The parameter constraints are robust to the inclusion of higher-order statistics. 
}
\label{fig:ld_b0}
\end{figure}

\noindent Throughout this work, we have omitted the galaxy bispectrum when performing cosmological analyses using simulation-based priors. In this Appendix, we apply the SBPs to the combined power spectrum and bispectrum dataset to test this assumption (noting that our SBPs only constrain EFT parameters appearing in the power spectrum). 
Fig.~\ref{fig:ld_b0} compares the cosmological constraints from the $P_\ell$ and $P_\ell+B_0$ analyses, both adopting SBPs. 

We find that including the galaxy bispectrum does not appreciably affect our $\ld$ parameter constraints. 
This validates our analysis pipeline and motivates dropping the galaxy bispectrum when using SBPs.
Due to the similarity in the posteriors, we do not report the
marginalized credible intervals for the $P_\ell+B_0$ analyses in this work.

Turning to the dynamical dark energy analysis, Tab.~\ref{tab:chi22} reports the $\chi^2_{\rm MAP}$ values for SBP analyses including the DESI DR1 bispectrum monopole.
When we do not include supernova distance measurements, the preference for evolving dark energy using SBPs is $2.5\sigma$, which is $0.3\sigma$ higher than in the analysis without the galaxy bispectrum. When the supernova dataset is included, adding the DESI DR1 bispectrum monopole reduces the preference for the $\wa$ model by $0.2\sigma$, leading to an overall significance of $1.2\sigma$. Overall, we conclude that the goodness-of-fit and corresponding preference for the $\wa$ model remain broadly stable upon the inclusion of higher-order statistics.

\begin{table*}[!htb]
    \centering
    %\resizebox{\linewidth}{!}{
    \begin{tabular}{lcccccc}
    \toprule
    Model/Dataset 
    & $M_\nu\,{\rm [eV]}$ 
    & $\Omega_m$ 
    & $H_0$ 
    & $\sigma_8$ 
    & $w_0$ 
    & $w_a$
    \\
    \hline
    %\enspace
    $\bm{\Lambda}$\textbf{CDM+}$\bm{M_\nu}$ &  &  &  &  &  & \\
    $\bao+P_\ell+B_0$ (GSBP)
    & $\enspace <0.239\enspace$ 
& $\enspace 0.3015_{-0.0071}^{+0.0065}\enspace$ 
& $\enspace 68.85_{-0.61}^{+0.60}\enspace$ 
& $\enspace 0.734_{-0.023}^{+0.023}\enspace$ %temp
& $\enspace -\enspace$
& $\enspace -\enspace$
    \\
    $\bao+P_\ell+B_0+\cmb$ (GSBP)
    & $<0.0782$ 
& $0.2989_{-0.0036}^{+0.0035}$ 
& $68.62_{-0.29}^{+0.29}$ 
& $0.8105_{-0.0055}^{+0.0071}$ %temp 
& $-$
& $-$
    \\
    \hline
    $\bm{w_0w_a}$\textbf{CDM+$\bm{M_\nu}$} &  &  &  &  &  & \\
    $\cmb+\bao+P_\ell+B_0+\sn$ (GSBP)
    & $<0.124$ 
& $0.3092_{-0.0056}^{+0.0056}$ 
& $67.65_{-0.58}^{+0.57}$ 
& $0.8040_{-0.0087}^{+0.0089}$ 
& $\enspace -0.877_{-0.053}^{+0.053}\enspace $ 
& $\enspace -0.39_{-0.18}^{+0.21}\enspace$ %temp
    \\
    \hline
    \end{tabular}
    %}
    \caption{\textbf{Massive neutrinos with Gaussian SBPs}: Constraints on cosmological parameters where the sum of neutrino masses
is allowed to vary, when using approximate Gaussian simulation-based priors.
    \label{tab:mnu2}
    }
\end{table*}

\section{Sensitivity to quasar priors}
\label{app:qso}

\noindent \resub{In the main analysis, we employed state-of-the-art HOD models to generate the SBPs for all tracers except quasars, for which we adopted more restrictive HOD models (excluding assembly bias and baryonic feedback). In this Appendix, we explore the sensitivity of the results to the choice of priors for the QSO sample.}
%of our results to the choice of simulations-based priors for this tracer. 

\begin{figure}[!t]
	\centering
	\includegraphics[width=1\linewidth]{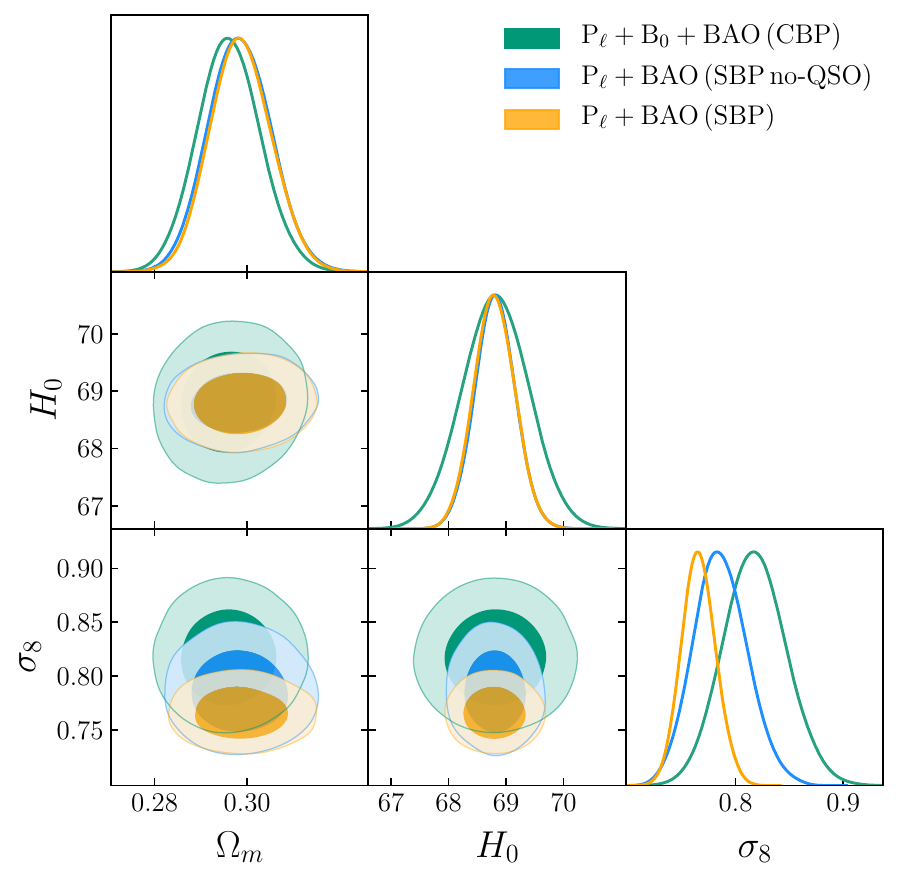}
	\caption{\resub{\textbf{SBP analysis with QSO conservative priors}: Posterior distributions for the parameters $\Omega$, $H_0$, $\sigma_8$ obtained from the following datasets: (1) $P_\ell$ + $B_0+\bao$ obtained using the conservative Bayesian priors (CBP); (2) $P_\ell+\bao$ analyzed with the simulation-based priors (SBP) used in the main analysis of this work; (3) $P_\ell+\bao$ analyzed with SBPs for all tracers except the QSO sample, for which conservative priors were adopted instead (SBP no-QSO). SBP analysis, which adopts conservative priors for QSOs, yields a $1\sigma$ higher value of $\sigma_8$, which is $20\%$ stronger than the baseline CBP result.}
    }
    \label{fig:qso}
\end{figure}

\resub{We perform the analysis using SBPs for all tracers except quasars, for which we adopt the conservative EFT priors from the baseline analysis.
Fig.~\ref{fig:qso} shows constraints on the cosmological parameters, with the one-dimensional marginalized constraints listed in Tab.~\ref{tab:qso}.}
\begin{table}[!t]
    \centering
    %\resizebox{\linewidth}{!}{
    \begin{tabular}{lccc}
    \toprule
    Dataset 
    %& $\omega_{cdm}$ 
    & $\Omega_m$ 
    & $H_0$ 
    & $\sigma_8$ 
    %& $S_8$  
    \\
    \hline
    %\enspace
    % SBP
    %\textbf{SBP} &  &  &  & & \\
%     $P_\ell$ (SBP)
%     & $\enspace 0.1210_{-0.0055}^{+0.0049}\enspace$
%     & $\enspace 0.292_{-0.011}^{+0.010}\enspace$ 
% & $\enspace 70.18_{-0.94}^{+0.86}\enspace$ 
% & $\enspace 0.766_{-0.016}^{+0.017}\enspace$ 
% & $\enspace 0.756_{-0.022}^{+0.020}\enspace$ %final2
%     \\
    $P_\ell\!+\!\bao$ (SBP no-QSO) %[QSO conservative]
    %& $0.1185_{-0.0037}^{+0.0033}$ 
& \multirow{1}{*}{$0.2984_{-0.0068}^{+0.0067}$}
& \multirow{1}{*}{$68.81_{-0.35}^{+0.35}$} 
& \multirow{1}{*}{$0.786_{-0.026}^{+0.024}$}
%& $0.764_{-0.018}^{+0.018}$ %final2
    \\
% {[except QSO]}
% & 
% &  
% & 
%     \\
    $P_\ell\!+\!\bao$ (SBP)
    %& $0.1185_{-0.0037}^{+0.0033}$ 
& $0.2987_{-0.0066}^{+0.0066}$ 
& $68.80_{-0.35}^{+0.35}$ 
& $0.766_{-0.016}^{+0.015}$ 
%& $0.764_{-0.018}^{+0.018}$ %final2
    \\\hline
%     $P_\ell+\bao+\cmb$ (SBP)
%     & $0.1175_{-0.0006}^{+0.0006}$ 
% & $0.2999_{-0.0036}^{+0.0033}$ 
% & $68.49_{-0.28}^{+0.28}$ 
% & $0.8057_{-0.0049}^{+0.0049}$ 
% & $0.8055_{-0.0069}^{+0.0070}$ %final2
%     \\\hline
%     % CBP
%     \textbf{CBP} &  &  &  & \\
%     $P_\ell$ 
%     & $0.1122_{-0.0067}^{+0.0068}$ 
% & $0.274_{-0.013}^{+0.012}$ 
% & $70.22_{-1.06}^{+1.06}$ 
% & $0.825_{-0.033}^{+0.033}$ 
% & $0.787_{-0.036}^{+0.036}$ %final2
%     \\
%     $P_\ell+B_0$ 
%     & $0.1189_{-0.0065}^{+0.0055}$ 
% & $0.284_{-0.012}^{+0.010}$ 
% & $70.67_{-1.05}^{+1.05}$ 
% & $0.811_{-0.031}^{+0.028}$ 
% & $0.789_{-0.035}^{+0.032}$ %final2
%     \\
    $P_\ell\!+\!B_0\!+\!\bao$ (CBP)
    %& $0.1174_{-0.0039}^{+0.0039}$ 
& $0.2961_{-0.0067}^{+0.0067}$ 
& $68.82_{-0.58}^{+0.58}$ 
& $0.818_{-0.029}^{+0.029}$ 
%& $0.813_{-0.031}^{+0.031}$ %final2
    \\
%     $P_\ell+B_0+\bao+\cmb$ 
%     & $0.1172_{-0.0006}^{+0.0006}$ 
% & $0.2983_{-0.0034}^{+0.0035}$ 
% & $68.62_{-0.28}^{+0.28}$ 
% & $0.8091_{-0.0056}^{+0.0051}$ 
% & $0.8068_{-0.0074}^{+0.0075}$ %final2
%     \\\hline
%     $\cmb$ 
%     & $0.1202_{-0.0012}^{+0.0012}$ 
% & $0.3164_{-0.0073}^{+0.0072}$ 
% & $67.28_{-0.53}^{+0.53}$ 
% & $0.8123_{-0.0051}^{+0.0052}$ 
% & $0.834_{-0.012}^{+0.012}$ %final2
%     \\
    \hline
    \end{tabular}
    %}
    \caption{\resub{\textbf{SBP analysis with QSO conservative priors}: Mean and 68\% confidence intervals on $\ld$ cosmological parameters obtained from various combinations of datasets, adopting conservative Bayesian priors (CBP), simulation-based priors (SBP) used in the main analyses, and simulation-based priors for all tracers except the QSO sample, for which conservative priors were adopted instead (SBP no-QSO). Two-dimensional posteriors are shown in Fig.~\ref{fig:qso}.
    }}
    \label{tab:qso}
\end{table}

\resub{We find that adopting conservative priors for the QSO power spectrum shifts the mean value of $\sigma_8$ upward by approximately $1\sigma$ and increases the error-bar by nearly $30\%$ relative to the SBP analysis presented in the main analysis of this work. As a result, the shift with respect to the baseline CBP result is reduced from $1.8\sigma_{\rm CBP}$ to $1.2\sigma_{\rm CBP}$, improving the consistency with the standard EFT-based analysis.}

\resub{Our results indicate that the preference for a low $\sigma_8$ in the SBP analysis is driven by the more restrictive HOD models used to generate the SBPs for the quasar sample.
Note that the quasar sample has a significantly higher redshift (and a large effective volume) and therefore is more sensitive to the initial conditions, which encode the bulk of cosmological information.
It is thus important to accurately model the SBPs for this tracer, as they represent an important systematical uncertainty in our analysis. 
In future work, we plan to derive the SBPs for quasars using more general galaxy formation models to determine whether the $\sigma_8$ shift reported in the main analysis of this work represents a genuine physical effect or it is merely driven by the simplified assumptions used to generate the SBPs for this tracer.
% tracer has a lot higher $z$ and thus has a higher sensitivity to the initial conditions, encoding the main cosmological dependency 
% It is important to derive SBPs for quasars using more general galaxy formation models.
% We plan to investigate this in future work.
}

\section{Analysis with Gaussian simulation-based priors}
\label{app:Gauss}

\noindent \cite{Chen:2025jnr} presented an alternative 
modeling technique for the EFT parameter distribution based on Gaussian mixture models, instead of the normalizing flow architecture used in the main text. 
For completeness, in this Appendix we present results based 
on the Gaussian mixture model with a single component, \textit{i.e.}\ we use a Gaussian approximation to EFT parameter distribution which match the mean and covariances of the EFT parameter distributions to those from HOD realizations \citep[cf.,][]{Akitsu:2024lyt}, but 
ignore higher order moments. We refer to this approximation
as Gaussian SBPs (GSBPs).

Key constraints on the 
$w_0w_a$CDM model and the neutrino masses 
are presented in Tabs.\,\ref{tab:w0wa2}\,\&\,\ref{tab:mnu2}.
While these results are broadly consistent with those from our fiducial analysis, we find that the use of GSBPs leads to weaker bounds than the normalizing flow SBPs. This underlines the importance of a robust modeling of the EFT priors, in order to accurately capture non-Gaussian correlations between EFT parameters. The inclusion of additional Gaussian mixture model modes may improve the agreement between the two types of SBP; we defer this test to future work.

\bibliography{short.bib}

\end{document}